\documentclass[aps,amsfonts,pra,twocolumn,showpacs]{revtex4}
\usepackage{epsfig,amsmath,amssymb,bm,epsf,graphicx,psfrag}
\usepackage[all]{xy}
\usepackage{graphicx}
\usepackage{color}

\pdfoutput=1

\newcommand{\ncd}{\newcommand}
\ncd{\RMbra}{\langle {\cal{R}}(r,m)|}
\ncd{\RMket}{|{\cal{R}}(r,m)\rangle}

\DeclareMathOperator{\Tr}{Tr}
\DeclareMathOperator{\supp}{supp}

\newcommand{\R}{\mathbb{R}}

\newcommand{\Z}{\mathbb{Z}}

\newcommand{\vphi}{\varphi}
\DeclareMathOperator{\GL}{GL}
\DeclareMathOperator{\SL}{SL}
\DeclareMathOperator{\AGL}{AGL}
\DeclareMathOperator{\Sp}{Sp}
\DeclareMathOperator{\ASp}{ASp}
\DeclareMathOperator{\Ker}{Ker}
\DeclareMathOperator{\im}{Im}

\newcommand\ket[1]{|#1\rangle}
\newcommand\bra[1]{\langle #1|}
\newcommand\braket[2]{\langle #1|#2 \rangle}
\newcommand\ketbra[2]{|#1\rangle \langle #2|}
\newcommand\proj[1]{|#1\rangle \langle #1|}

\newtheorem{Theorem}{Theorem}

\newtheorem{Cor}{Corollary}

\newtheorem{Lemma}{Lemma} 

\newtheorem{Def}{Definition}

\newtheorem{Crit}{Criterion}

\begin{document}

\title{Wigner function negativity and contextuality in quantum computation on rebits}

\author{$\text{Nicolas Delfosse}^1$, $\text{Philippe Allard Guerin}^2$, $\text{Jacob Bian}^2$,  $\text{Robert Raussendorf}^2$}
\affiliation{1: D{\'e}partment de Physique, Universit{\'e} de Sherbrooke, Sherbrooke, Qu{\'e}bec, J1K 2R1, Canada,\\ 2: Department of Physics and Astronomy, University of British
Columbia, Vancouver, British Columbia V6T 1Z1, Canada}
\date{\today}

\begin{abstract}
We describe a universal scheme of quantum computation by state injection on rebits (states with real density matrices). For this scheme, we establish contextuality and Wigner function negativity as computational resources, extending results of [M. Howard {\em{et al.}}, Nature 510, 351--355 (2014)] to two-level systems. For this purpose, we define a Wigner function suited to systems of $n$ rebits, and prove a corresponding discrete Hudson's theorem. We introduce contextuality witnesses for rebit states, and discuss the compatibility of our result with  state-independent contextuality.
\end{abstract}

\pacs{03.67.Mn, 03.65.Ud, 03.67.Ac}

\maketitle

\section{Introduction}

In quantum computation by state injection (QCSI) \cite{Magic}, the set of quantum gates is by construction not universal. This restriction is made up for by the injection of states that could not be created within the scheme itself, the so-called magic states. 

Besides its promise for the realization of fault-tolerant quantum computation, QCSI is of fundamental theoretical interest. Since the magic states enable universality, one is led to ask: {\em{Precisely which quantum properties of these states are responsible for the gain in computational power?}}

Contextuality \cite{Bell}-\cite{Acin} and negativity of Wigner functions have recently been proposed as the quintessential quantum properties of magic states; See \cite{Howard},\cite{Galv1},\cite{Veitch}. Con\-tex\-tuality is an obstruction to modelling the inherent randomness of quantum measurement in a statistical mechanics fashion, namely by a probability distribution over configurations with predetermined measurement outcomes for all measurable observables. Wigner functions \cite{Wigner}-\cite{Gross} are the closest quantum analogue of probability distributions over phase space. The  key difference is that Wigner functions can assume negative values, and this negativity is taken as an indication of quantumness. Despite their separate origins in the fields of quantum optics and foundations of quantum mechanics,  Wigner function negativity and contextuality are closely related indicators of non-classical behaviour \cite{Spekkens}, \cite{Howard}. 

The reason for the appearance of Wigner functions in the discussion of QCSI is their relation \cite{Galv1}, \cite{Gross} to the stabilizer formalism \cite{DG}.  The stabilizer formalism  is also relevant for QCSI, since the restricted gate set therein is typically chosen to be the Clifford gates.  These gates are indeed not universal, and---if supplemented only with Pauli measurements and stabilizer states---can be efficiently classically simulated by stabilizer techniques. 

An epitome for the link between Wigner functions and QCSI via the stabilizer formalism is the discrete Hudson's theorem \cite{Gross}, which says that in Hilbert spaces of odd prime-power (hence finite) dimension, the pure states with positive Wigner function are exactly the stabilizer states. Thus, stabilizer states are ``classical'' from both the perspectives of Wigner functions and QCSI.  In the wake of this result, contextuality and Wigner function negativity have been established as quantum resources for QCSI with qudits of odd prime dimension \cite{Veitch}, \cite{Howard}. 

Extending these properties to 2-level systems is pertinent, since quantum algorithms are typically formulated in terms of qubits. But attempts to do so hit barriers: As for the Wigner functions, many constructions cannot be adapted to qubits \cite{Gross}, \cite{Mari}; and for the remaining ones, \cite{Galv1}, \cite{Galv2}, the discrete Hudson's theorem breaks down. There are qubit stabilizer states with negative Wigner function. As for contextuality, it  now arises in its state-independent form \cite{Merm}. In result, every quantum state of more than one qubit can be considered contextual \cite{Howard}, which is at odds with viewing contextuality as a resource possessed only by special states.

Here, we establish Wigner function negativity and contextuality as necessary resources for QCSI on two-level systems. We achieve this at the price of restricting from qubits to rebits, i.e., real density matrices of $n$ two-level systems. This restriction does not affect universality \cite{rebit_rudolph}. The role that was previously played by the stabilizer states is now played by the CSS-states \cite{CSS}, and the group of Clifford gates is replaced by the subgroup of CSS-ness preserving Clifford gates. Within this new setting, we resurrect a discrete Hudson's theorem, as well as a number of related properties of the Wigner function. Furthermore, the restriction to CSS-ness preserving operations permits us to carve out a computational scheme of rebit QCSI that is free of state-independent contextuality, even if this phenomenon exists in rebits.\smallskip 

This paper is organized as follows. Section~\ref{QCSI} summarizes the known results on the roles of  contextuality and negativity in qudit QCSI, and defines our setting for rebits. In Section~\ref{reQCSI} we present a universal scheme of quantum computation by state injection on rebits. In Section~\ref{Wigner}, we construct a matching Wigner function, equipped with a discrete Hudson's theorem and extended Gottesman-Knill theorem. In Section~\ref{Cont} we provide necessary and sufficient conditions for contextuality in terms of the Wigner function. Section~\ref{CNC} contains our results on contextuality and negativity as resources in rebit-QCSI. We conclude in Section~\ref{Concl}.

\section{Quantum computation by state injection}
\label{QCSI}

\medskip

QCSI has four operational quantum components: the restricted unitary gates, the restricted measurements, the cheap states and the magic states. The cheap states are those that can be produced from  sequences of measurements from the restricted set and restricted unitary gates, possibly classically conditioned on measurement outcomes. The classical side-processing is unrestricted.

A typical choice for the restricted operations is that they live within the stabilizer world. That is, the restricted set of unitary gates is in the group of Clifford gates, the restricted set of observables is the Pauli observables or a subset thereof, and the cheap states are stabilizer states. 

\subsection{Summary of the qudit case}
\label{SumDit}

For the case of odd prime local dimension $d$, QCSI has been investigated for the restricted gate set being the Clifford gates  \cite{Veitch},\cite{Howard}. For this scenario, two essential quantum  properties of the magic states have been identified, namely the negativity of their Wigner function, and their contextuality with respect to stabilizer measurements. Specifically, it has been established that
\begin{itemize}
\item[(i)]{Negativity in the Wigner function of raw magic states is necessary for successful magic state distillation (Theorem 3 in \cite{Veitch}) and for the hardness of classical simulation of QCSI (Theorem~1 in \cite{Veitch}).}
\item[(ii)]{Contextuality of magic states w.r.t stabilizer measurements is necessary for universality of QCSI \cite{Howard}.}
\end{itemize}
The Wigner function plays a dual role for QCSI. It is relevant for the phenomenology observed (see above), but it is also deeply involved in the mathematical description of the computational scheme.  This is revealed in the following five properties, which hold for odd prime $d$ when the restricted operations belong to the stabilizer world,
\begin{itemize}
\item[(iii)]{The set of stabilizer states is singled out by a Hudson's theorem as the set of pure states with non-negative Wigner function \cite{Gross}.}
\item[(iv)]{The set of Clifford gates is singled out as the set of unitaries that transform the Wigner function covariantly \cite{Gross}.}
\item[(v)]{Clifford gates and stabilizer measurements preserve positivity of the Wigner function \cite{Veitch}.}
\item[(vi)]{Necessary and sufficient conditions for contextuality w.r.t. the restricted set of measurements can be expressed in terms of the Wigner function \cite{Howard}.}
\item[(vii)]{For one-qudit states, negativity of the Wigner function and contextuality w.r.t. measurements from the restricted set are the same \cite{Howard}.} 
\end{itemize}
The above physical properties (i) and (ii) are consequences of the structural properties (iii) - (vii). For example, an efficient classical simulation method  for the evolution of states with non-negative Wigner function under the restricted gates can be built on properties (iv) and (v) \cite{Veitch}. Its existence directly implies (i). Furthermore, Hudson's theorem (iii) connects this simulation method with the Gottesman-Knill theorem.

\subsection{Trouble with qubits}

For systems of qubits, both the employed contextuality witnesses \cite{CSW} and Wigner functions run into difficulty. As for contextuality, if the goal is to establish it as a quantum resource, one has to overcome a problem posed by the phenomenon of state-independent contextuality which is revealed, for example, by the Mermin square and star \cite{Merm}. Mermin's square can be translated into a contextuality witness for which {\em{all}} quantum states of $n\geq 2$ qubits come out contextual \cite{Howard}. If contextuality is generic then it cannot be a resource.

As for the Wigner functions, many of the Wigner functions proposed for Hilbert spaces of finite dimension $d^n$ require for their definition the existence of $2^{-1}$ in $\mathbb{F}_d$, and thus do not apply to the qubit case $d=2$; for examples see e.g. \cite{Gross}, \cite{Mari}. 

Yet some Wigner functions do survive the transition to $d=2$; see e.g. \cite{Galv1}, \cite{Galv2}. However, in these cases, the general connection with the stabilizer world breaks down. Not all stabilizer states have non-negative Wigner function anymore, and the Wigner function no longer transforms covariantly under all Clifford operations.\medskip

For the construction \cite{Galv1}, \cite{Galv2},  Wigner function negativity of the magic states is necessary for universality. Therein, not a single Wigner function is considered but instead the whole class introduced in \cite{Gibbons}. A ``classical'' state must be positively represented for each of these Wigner functions. The number of pure  $n$-qubit states for which this holds is super-exponentially small compared to the number of $n$-qubit stabilizer states \cite{Gross}.

\subsection{Rebits}\label{Reb}

In this paper, we discuss the case of local dimension $d=2$. We present a universal scheme of QCSI for which contextuality and Wigner function negativity are established as necessary quantum resources.  The price we pay is that we have to restrict from qubits to rebits. Specifically, we require that the density matrix of the processed quantum state $\rho$ is real; i.e. at each point in the quantum computation it holds that
$$
\langle x |\rho | y \rangle \in \mathbb{R}, 
$$ 
for all $|x\rangle$, $|y \rangle$  in the computational basis.

For the discussed scheme of rebit QCSI, the set of cheap states is the CSS states, the set of restricted gates is the CSS-ness preserving Clifford gates, and the allowed measurements are of observables from the set
\begin{equation}
{\cal{O}} = \{X(\textbf{a}_X), Z(\textbf{a}_Z)| \textbf{a}_X,\textbf{a}_Z \in \mathbb{Z}_2^n \}.
\end{equation} 
That is, in our construction the restricted operations belong to the CSS-stabilizer world, rather than the more general stabilizer world.

\subsection{CSS states and CSS-ness preserving Clifford operations}\label{CSS}

Calderbank-Shor-Steane (CSS) states are a subset of the stabilizer states. They are defined by the property that for any CSS state $|\psi\rangle$, the corresponding Pauli stabilizer group ${\cal{S}}(|\psi\rangle)$ decomposes into an $X$- and a $Z$-part; i.e., ${\cal{S}}(|\psi\rangle) = {\cal{S}}_X(|\psi\rangle)\times {\cal{S}}_Z(|\psi\rangle)$, where all elements of ${\cal{S}}_X(|\psi\rangle)$ and ${\cal{S}}_Z(|\psi\rangle)$ are of the form $X(\textbf{a}_X)$ and $Z(\textbf{a}_Z)$, respectively. All CSS states are real, but not all real stabilizer states are of CSS type.

We now characterize the CSS-ness preserving transformations. Denote by $\Omega$ the set of pure CSS-states and by $G_{CSS}$ the subgroup of the $n$-qubit Clifford group $C_n$ which preserves the set $\Omega$ of CSS states,
\begin{equation}\label{GcssDef}
G_{CSS}= \{ g\in C_n|\, g|\Psi\rangle \in \Omega,\, \forall |\Psi\rangle \in \Omega\}.
\end{equation}
The following can be said about the structure of $G_{CSS}$.
\begin{Lemma}\label{appcor:G_CSS_generators}
The $n$-rebit CSS-ness preserving group $G_{CSS}$ is 
\begin{equation} \label{appeqn:CSS_generators}
G_{CSS}=\left\langle \bigotimes_{i=1}^n H_i, \text{CNOT}(i, j), X_i, Z_i \right\rangle,
\end{equation}
where $i, j \in \{1, 2,  \dots, n\}$ and $i \neq j$. 
We have the group isomorphism
\begin{equation} \label{appeqn:CSS_structure}
G_{CSS} / \{\pm I\} = \Z_2^{2n} \rtimes \left( \GL_n(\Z_2) \rtimes \Z_2 \right).
\end{equation}
\end{Lemma}
In Eq.(\ref{appeqn:CSS_structure}), the component $\Z_2^{2n}$ corresponds to the Pauli operators $T_{\bf u}$, the component $\GL_n(\Z_2)$ corresponds to the group generated by the $CNOT$, and the subgroup $\Z_2$ is generated by the simultaneous Hadamard gate $\otimes_i H_i$.

Since the set 
${\cal{O}} = \{Z({\bf u}) \ | \ {\bf u} \in \Z_2^n\} \cup \{X({\bf v}) \ | \ {\bf v} \in \Z_2^n\}$ is mapped onto itself by conjugation under gates from the group on the r.h.s. of Eq.~(\ref{appeqn:CSS_generators}), it is clear that this group is a subgroup of $G_{CSS}$ as defined in Eq.~(\ref{GcssDef}). That it is indeed all of $G_{CSS}$ is proved in Appendix~\ref{appsection:CSS_Clifford}.

\medskip
The set $\Omega$ of ``cheap'' CSS states, the CSS-ness preserving unitary gates $G_{CSS}$ and the projective measurements of observables in ${\cal{O}}$ form a compatible classical reference structure for QCSI, in the sense that none of these operations can map states inside $\Omega$ to states outside $\Omega$.

\section{Universal quantum computation by state injection on rebits}
\label{reQCSI}

It has been shown in  \cite{rebit_rudolph} that rebits are sufficient for universal quantum computation. In that scheme, first, a quantum state of $n$ qubits,
$$
|\psi\rangle  = \sum _{\mathbf{v} \in \mathbb{Z}_2 ^n} r_\mathbf{v} e^{i \theta_\mathbf{v}}|\mathbf{v}\rangle,
$$
is encoded into a state of  $n+1$ rebits,
\begin{equation}
\overline{|\psi\rangle} =\sum _{\mathbf{v} \in \mathbb{Z}_2 ^n} \left( r_\mathbf{v} \cos \theta_\mathbf{v} |\mathbf{v}\rangle\otimes  |R\rangle +  r_\mathbf{v} \sin \theta_\mathbf{v} |\mathbf{v}\rangle \otimes |I\rangle  \right).
\label{rebit_E}
\end{equation}
The additional rebit, with basis states $|R\rangle = |0\rangle$ and $|I\rangle = |1\rangle$, allows to keep track of the real and imaginary parts of the unencoded $n$-qubit state. 
Second, an encoded set of gates is constructed which (i) is universal, and (ii) preserves real-ness of the states in Eq.~(\ref{rebit_E}).\medskip

Using the encoding Eq.~(\ref{rebit_E}), we construct a universal scheme of QCSI on rebits. The restricted gate set  therein consists of CNOT-gates, the simultaneous Hadamard-gate $H_\text{all}:=\bigotimes_{i=1}^n H_i$, and Pauli-flips $X_i$, $Z_j$;
i.e., 
$$
\langle {\cal{G}}_\text{restricted}\rangle = G_{CSS}.
$$
These unitary gates are supplemented by measurements of observables in the set ${\cal{O}}$, or, w.l.o.g., of observables $\{Z_i|\,i=1,..,n\}$.

The (unitary) Pauli operators and the simultaneous Hadamard-gate can be dispensed with, because they can be propagated past the readout measurements. This is a consequence of the well-known propagation relations for Pauli operators under conjugation by Clifford gates, and $CNOT(i,j) H_\text{all} = H_\text{all} CNOT(j,i)$. If those gates are eliminated, we remain with the CNOT-gates and measurements of $X_i$ and $Z_i$. We note that this is precisely the set of gates which can be performed fault-tolerantly on the surface code \cite{Kit1} using defect braiding \cite{RH07}. However, for the  present purpose, we keep the redundant $H_\text{all}$ and Pauli flips in the restricted gate set.

For the universal gate set, we pick 
$${\cal{G}}_\text{universal}=\left\{CNOT(i,j), H_i, \exp(i\pi/8 \, Z_i)\right\},$$
supplemented with measurements of the Pauli observables $Z_i$, for $i=1,..,n$.
 
We now demonstrate that the encoded versions of these gates can be realized only using the gates from the restricted set and the injection of two types of ancilla states, $|A\rangle$ and $|B\rangle$, defined as 
\begin{equation}\label{and}
\begin{array}{rcl}
|A\rangle &= &\displaystyle{\frac{|0\rangle |R\rangle + \cos \frac{\pi}{4} |1\rangle |R\rangle +\sin \frac{\pi}{4} |1\rangle | I \rangle}{\sqrt{2}},}\\
|B\rangle &=& \displaystyle{\frac{|0\rangle |+\rangle + |1\rangle |-\rangle}{\sqrt{2}}.}\\
\end{array}
\end{equation}
The ancilla $|A\rangle$ is the encoded $(|0\rangle + e^{i\pi/4}|1\rangle)/\sqrt{2}$, with respect to the encoding of Eq.~(\ref{rebit_E}).\medskip

(a) {\em{The measurement of $Z_i$.}} Since the Pauli-operator $Z$ is real, its measurement does not differentiate between the real and imaginary parts of the measured state, and $\overline{Z}_i \equiv Z_i$. Graphically,
$$
\parbox{5cm}{\includegraphics[width=5cm]{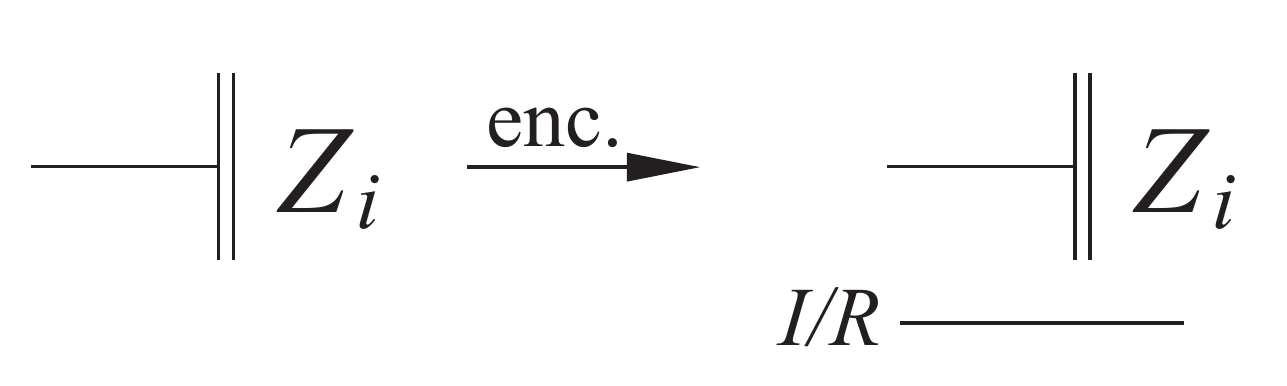}}.
$$

(b) {\em{The CNOT-gate between qubits $i$ and $j$.}} The CNOT-gate is real and hence does not mix the real and imaginary parts of the state it is applied to. Hence, $\overline{CNOT(i,j)} = CNOT(i,j)$.  Graphically,
$$
\parbox{4.5cm}{\includegraphics[width=4.5cm]{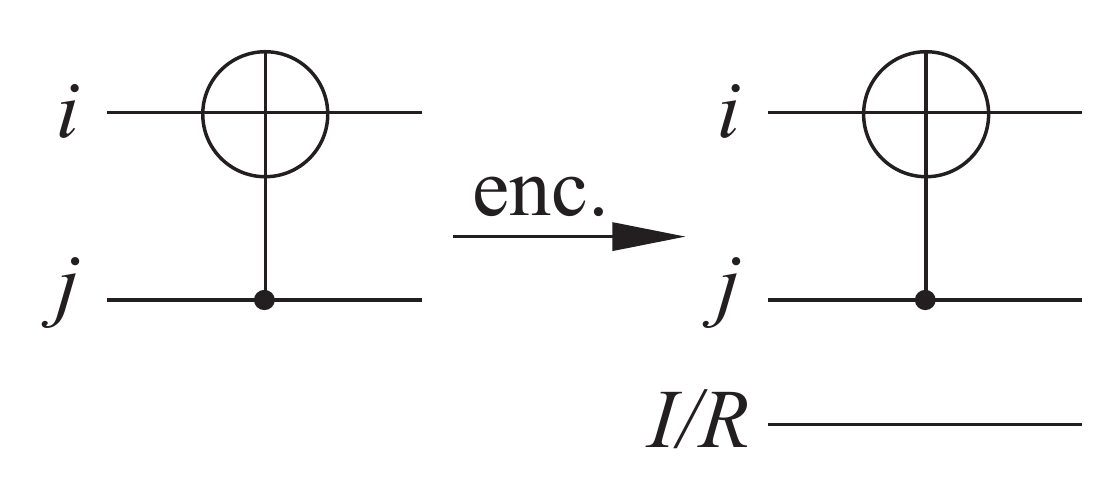}}.
$$

(c) {\em{The Hadamard gate $H_i$.}} The encoded Hadamard gate is realized by injection of an ancilla $|B\rangle$
into the circuit
$$
\parbox{7cm}{\includegraphics[width=5.5cm]{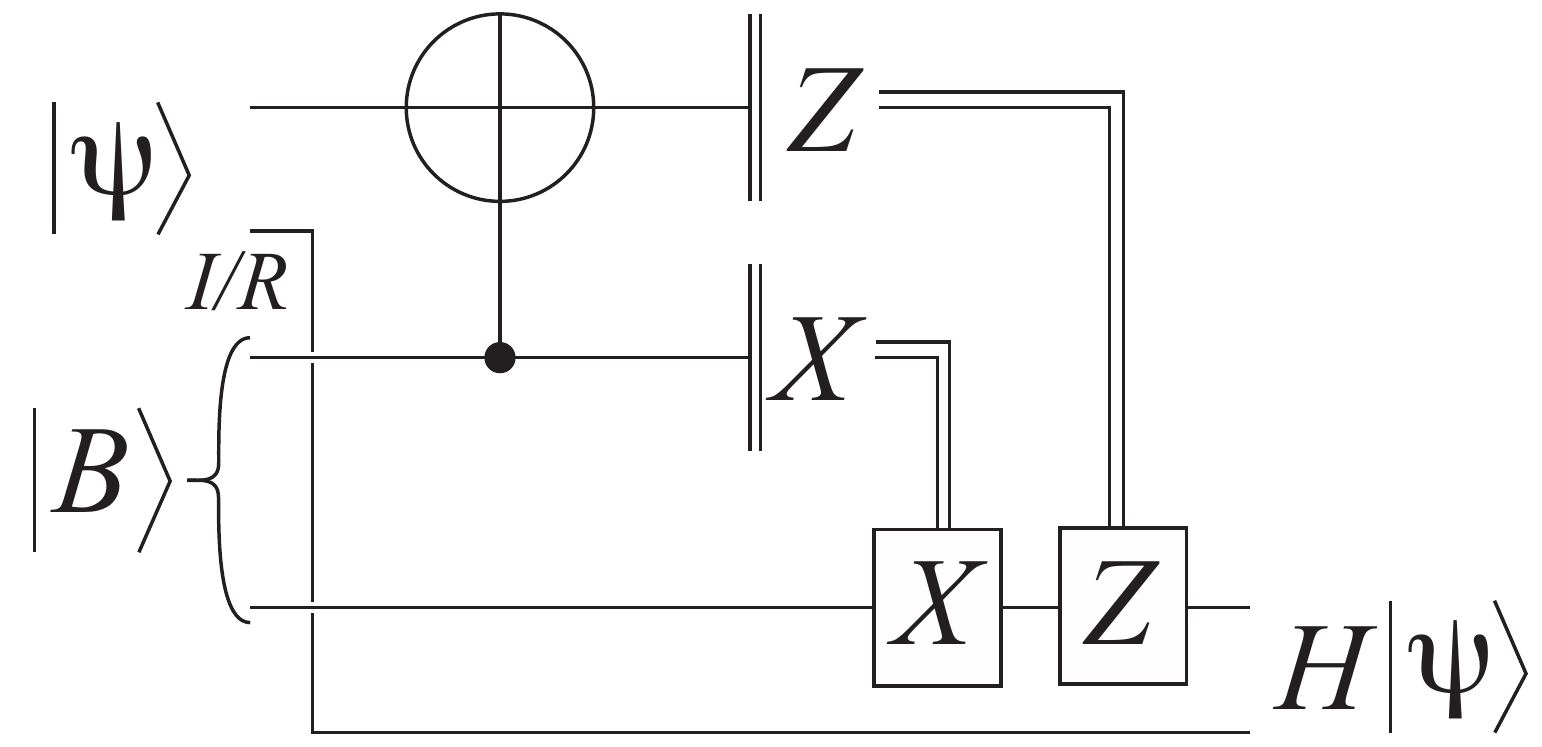}}.
$$

(d) {\em{The gate $\exp(i\pi/8 \, Z_i)$.}} The encoded version of this gate uses an ancilla states $|A\rangle$ and $|B\rangle$, and proceeds in two steps. The first step is a pre-processing jointly for all the $\exp(i\pi/8 \, Z_i)$ gates in the circuit. Namely, at the beginning of the computation, each ancilla state $|A\rangle$ is in its own separate code block. In the pre-processing step, all data and ancila rebits are merged into the same code block. The merging can be done two blocks at a time, and the corresponding circuit is 
$$
\parbox{6cm}{\includegraphics[width=6cm]{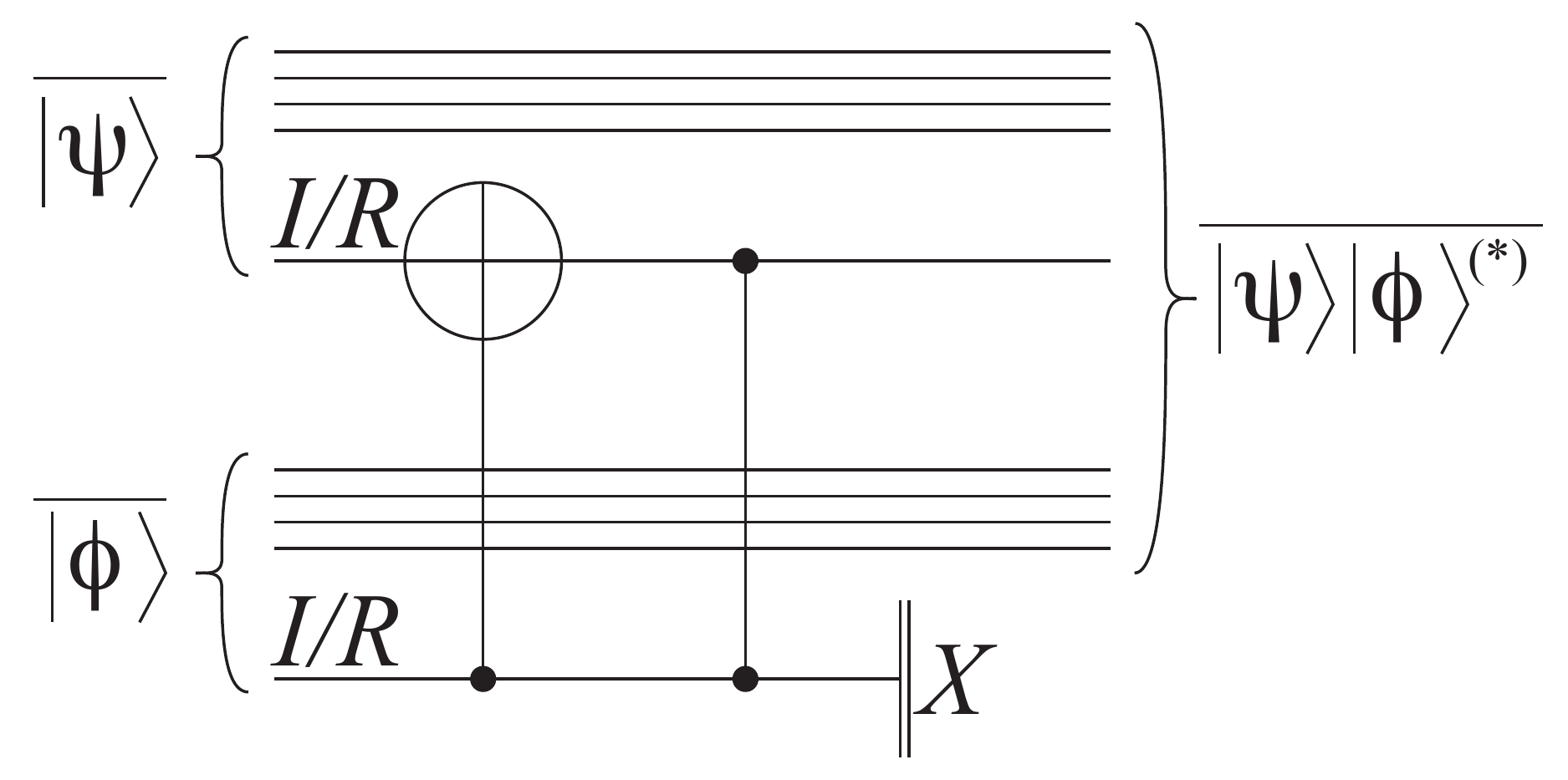}}.
$$
For a pair of encoded input states $\overline{|\psi\rangle}$, $\overline{|\phi\rangle}$, the result of the code merging circuit is $\overline{|\psi\rangle\otimes |\phi\rangle}$ or $\overline{|\psi\rangle\otimes |\phi\rangle^*}$, depending on the outcome of the $X$-measurement ($ |\phi\rangle^*$ denotes the state obtained from $|\psi\rangle$ by complex conjugation w.r.t. the computational basis). 

We will only ever use the code merging circuit for encoding the ancilla $|A\rangle = \overline{|\pi/8\rangle}$ into a single code block. Since $|\pi/8\rangle$ and $|\pi/8\rangle^*=|-\pi/8\rangle$ allow to perform the $\pi/8$-phase gate with the same efficiency, the probabilistic nature of the code merging circuit does not affect the computation.

The code merging circuit contains a conditional phase gate which is not part of the restricted gate set. It is realized via the following state-injection circuit,
$$
\parbox{6.5cm}{\includegraphics[width=6.5cm]{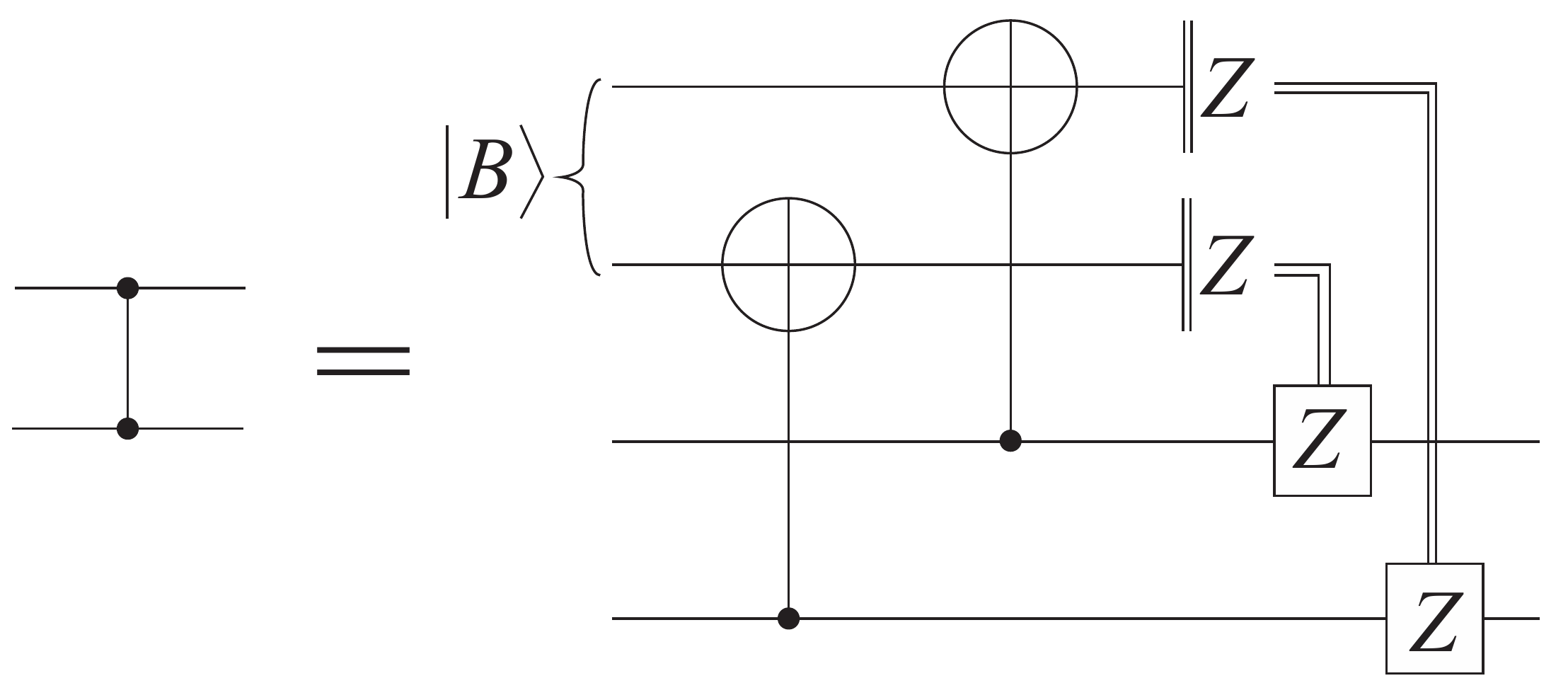}}.
$$
The second step then is the encoded version of the  standard state injection circuit for the $\pi/8$-gate \cite{NC},
\begin{equation}\label{circuitSI}
\parbox{3cm}{\includegraphics[width=3cm]{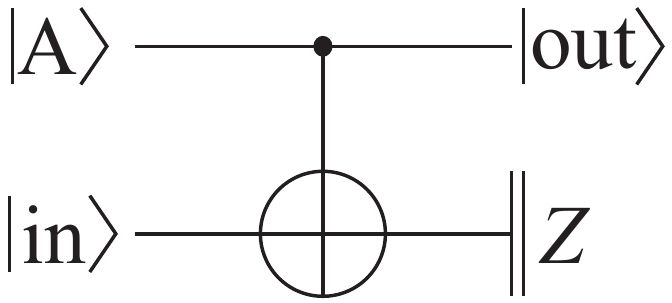}}
\end{equation}
This circuit consists solely of operations whose encoded versions we have already demonstrated.

\section{A Wigner function for rebits}
\label{Wigner}

In the last section we described a universal scheme of quantum computation by state injection on rebits, and here we construct the matching Wigner function. We first  propose the rebit Wigner function and examine its basic properties. Second, we prove a discrete Hudson's theorem for rebits. Third, we prove covariance of the rebit Wigner function under CSS-ness preserving Clifford unitaries; and finally show that the evolution of states with positive Wigner function under CSS-ness preserving Clifford unitaries and measurements can be efficiently classically simulated.

\subsection{Definition of a Wigner function for rebits}
\label{Widef}

We now proceed to construct the Wigner $W$ function for $n$-rebit states, which is suited to describe the computational scheme introduced in the previous section. It is a modification of the Wigner function $\tilde{W}$ \cite{Veitch}, \cite{Howard} for qubits. Of the properties (i) - (vii) listed in Section~\ref{SumDit} for the Wigner function on qudits, our rebit Wigner function has counterparts for properties (i) - (vi) but not for (vii).

In the qubit case, there are $4^n$ Pauli operators $T_a$,
\begin{equation}\label{T}
T_{(\textbf{a}_Z,\textbf{a}_X)} = Z(\textbf{a}_Z)X(\textbf{a}_X) ,\;\; \text{where}\;  \textbf{a}_Z,\textbf{a}_X \in \mathbb{Z}_2^n.
\end{equation}
Therein, $Z(\textbf{a}) = Z_1^{a_1} \otimes Z_2^{a_2}\otimes .. \otimes Z_n^{a_n}$, for all $(\textbf{a}_X,\textbf{a}_Z) \in \mathbb{Z}_2^n \times \mathbb{Z}_2^n $. We denote 
$$
\begin{array}{rcl}
V&:=& \{(\textbf{a}_X,\textbf{a}_Z)|\, \textbf{a}_X,\textbf{a}_Z \in \mathbb{Z}_2^n\} \cong \mathbb{Z}_2^{2n},\\
{\cal{T}} &:=& \{T_\textbf{a}|\, \textbf{a} \in V\}.
\end{array}
$$
The Pauli operators ${\cal{T}}$ form an orthonormal basis of the vector space of square matrices of size $2^n$ with complex coefficients endowed with the inner product defined by $(A, B) = \frac{1}{2^n} \Tr(A^\dagger B)$, 
\begin{equation}
\label{tr}
\text{Tr}(T_\textbf{a}^\dagger T_\textbf{b}) = 2^n \delta_{\textbf{a},\textbf{b}},\;\; \forall \; T_\textbf{a}, T_\textbf{b} \in \cal T.
\end{equation}
In the present work, we are interested in rebits, which are defined by symmetric real density operators. We consider the set
\begin{equation}\label{CalA}
\mathcal{A} := \{T_{\bf a} \ | \ ({\bf a}_Z, {\bf a}_X)=0 \mod 2 \},
\end{equation} 
which is an orthonormal basis of the space of symmetric matrices (see Lemma~\ref{lemma:A_orthonormal} in Appendix~\ref{WiProp}), and define
\begin{equation}
\label{WFa}
W_\rho(\textbf{u}): = \frac{1}{2^n} \text{Tr} (A_\textbf{u} \rho),
\end{equation}
with
\begin{equation}
\label{mwf}
A_{\textbf{0}} = \frac{1}{2^n} \sum_{T_\textbf{a} \in \cal{A} }T_\textbf{a}, \;\;\text{and}\;\; A_\textbf{u} = T_\textbf{u} A_\textbf{0} T_\textbf{u}^\dagger.
\end{equation}
For later use, denote
$
V_{\cal{A}}=\{\textbf{a}  \in V| \, (\textbf{a}_z,\textbf{a}_x)\!\! \mod 2=0\}.
$
Note that the operator $A_{\bf u}$ can also be written 
\begin{equation} \label{eqn:A_u_second_form}
A_{\bf u} = \frac{1}{2^n} \sum_{T_\textbf{a} \in \cal{A} } (-1)^{[{\bf u}, {\bf a}]} T_{\bf a}
\end{equation}
where $[{\bf u}, {\bf v}] = ({\bf u}_Z, {\bf v}_X) + ({\bf v}_Z, {\bf u}_X)$ is the symplectic inner product in $\Z_2^{2n}$.

When considering real states, the family $(A_\textbf{u})_{\textbf{u} \in V}$ is not a basis of the space of symmetric matrices since it contains too many matrices. 
Nevertheless, in close analogy with the qudit Wigner function \cite{Gross}, \cite{Veitch}, the rebit Wigner function of Eq.~(\ref{WFa}) has the following properties (compare with \cite{Veitch2}):
\begin{enumerate}
\item{Any real density matrix $\rho$ satisfies 
$$\rho=\sum_\textbf{u}W_\rho(\textbf{u})A_\textbf{u}.$$ $W$ is thus informationally complete.} 

\item{$W$ transforms covariantly under the group of CSS-ness pre\-ser\-ving  Clifford transformations.}

\item{The CSS-states are the only pure states with non-negative $W$ (discrete Hudson's theorem).}

\item{For all real density matrices $\rho$, $\sigma$,
$$
W_{\rho \otimes \sigma} = W_\rho \cdot W_\sigma.
$$}
\item{The trace inner product is given as
\begin{equation}\label{Trip}
\text{Tr}(\rho \sigma) = 2^n \sum_{\textbf{u} \in V} W_\rho(\textbf{u}) W_\sigma(\textbf{u}).
\end{equation}}

\item{The phase point operators satisfy $\text{Tr} A_\textbf{u} = 1$. Thus, $\text{Tr} B = \sum_\textbf{u} W_B(\textbf{u})$ for any symmetric operator $B$.}
\end{enumerate}
Property 1 is proven in Lemma~\ref{lemma:wigner_invertible} in Appendix~\ref{WiProp}, Property 2 in Section~\ref{section:hudson}, and Property 3 in Section~\ref{section:covariance}. Property~4 and 5 are shown in Appendix~\ref{WiProp}. Property 6 is an immediate consequences of Property~1.

\subsection{A discrete Hudson's theorem for rebits} \label{section:hudson}

The original Hudson's theorem in infinite-dimensional Hilbert space \cite{Hudson} singles out the Gaussian states as the pure states with positive Wigner function. This result has a counterpart in finite, odd prime-power dimension. Namely, the pure states with positive Wigner function are the stabilizer states \cite{Gross}. In this way, a connection between Wigner functions and the discrete world of the stabilizer formalism is established. For no known Wigner function defined on multiple qubits, this result carries over (See \cite{WB}, however, for a single qubit). 

Here, for the Wigner function defined in the previous section, we find that for multiple rebits a discrete Hudson's theorem holds with the stabilizer states replaced by the more special CSS states.

\begin{Theorem}\label{theo:hudson}
A pure real state $|\psi\rangle$ has non-negative Wigner function $W_\psi$ if and only if it is a CSS state.
\end{Theorem}
Recall that a Wigner function $W_\rho$ for some density operator $\rho$ is said to be non-negative if $W_\rho({\bf u}) \geq 0$ for all ${\bf u} \in V$, and is said to be negative otherwise.

In order to prove this result we follow the strategy pursued by Gross for the qudit case \cite{Gross}. First, we determine the Wigner function of CSS states in Section~\ref{section:hudson_CSS}, proving that these Wigner functions are non-negative. Then, in Section~\ref{section:hudson_positivity}, we consider a pure state with non-negative Wigner function and we prove that this function is precisely the Wigner function of a CSS state. 
Finally, the fact that the Wigner function is informationally complete allows us to conclude the proof of Theorem~\ref{theo:hudson} in Section~\ref{section:hudson_proof}.

\subsubsection{Wigner function of CSS states} \label{section:hudson_CSS}

We start by computing the Wigner function of pure CSS states.

\begin{Lemma}\label{prop:computation_wigner_CSS}
The Wigner function of a pure CSS state $\ket \psi$ is of the form
$$
W_\psi = \frac{1}{2^n} \delta_{{\bf t} + V_S},
$$
where ${\bf t}$ is a vector of $\Z_2^{2n}$ and $V_S = N^\perp \times N$ for some subspace $N$ of $\Z_2^n$. Moreover, every such function $\frac{1}{2^n} \delta_{{\bf t} + V_S}$ is the Wigner function of a CSS state.
\end{Lemma}
In particular, the Wigner function of a pure CSS state is non-negative.

\medskip
{\em{Proof of Lemma~\ref{prop:computation_wigner_CSS}.}} 
Let $\ket \psi$ be a CSS state. Its stabilizer group ${\cal{S}}$ is generated by $r$ independent operators $(-1)^{\alpha_i} Z({\bf a}_i)$, for $1 \leq i \leq r$, and $n-r$ independent operators $(-1)^{\alpha_i} X({\bf b}_i)$, for $r+1 \leq i \leq n$.
Denote by $N$ the subspace of $\Z_2^{n}$ generated by the vectors ${\bf b}_i$, for $r+1 \leq i \leq n$, so that
its orthogonal complement is $N^\perp = \langle {\bf a}_1, {\bf a}_2, \dots, {\bf a}_n \rangle$.

The elements of ${\cal{S}}$ are thus of the form $(-1)^{\alpha(\bf v)} T_{\bf v}$, where ${\bf v} \in N^\perp \times N$. Moreover, we can easily check that the phase $(-1)^{\alpha(\bf v)}$ defines a character of $N^\perp \times N$. Since every such character can be written as ${\bf v} \mapsto (-1)^{[\bf t, v]}$, for some vector $\textbf{t} \in \Z_2^{2n}$, we have
$$
{\cal{S}} = \{ (-1)^{[\bf t, v]} T_{\bf v} \ | \ {\bf v} \in N^\perp \times N \}.
$$
Denote by $V_{\cal{S}}$ the subspace $N^\perp \times N$ of $\Z_2^{2n}$, then
$$
\proj \psi = \frac{1}{2^n} \sum_{{\bf v} \in V_S} (-1)^{[{\bf t}, {\bf v}]} T_{\bf v}.
$$
This, together with the definition Eq.(\ref{eqn:A_u_second_form}) of $A_{\bf u}$ leads to
\begin{align*}
W_\psi({\bf u}) & = \frac{1}{2^n} \Tr(A_{\bf u} \proj \psi)\\
& = \frac{1}{2^{3n}} \sum_{{\bf v} \in V_{\cal{S}}} \sum_{{\bf a} \in V_{\cal A}} (-1)^{[{\bf t, v}]} (-1)^{[{\bf u, a}]} \Tr(T_{\bf a} T_{\bf v})\\
& = \frac{1}{2^{2n}} \sum_{{\bf v} \in V_{\cal{S}}} \sum_{{\bf a} \in V_{\cal A}} (-1)^{[{\bf t, v}] + [{\bf u, a}]} \delta_{\bf a, v}\\
& = \frac{1}{2^{2n}} \sum_{{\bf v} \in V_{\cal{S}}} (-1)^{[{\bf v, t+u}]}\\
& = \frac{1}{2^n} \delta_{V_{\cal{S}}} ({\bf t+u})\\
& = \frac{1}{2^n} \delta_{{\bf t}+V_{\cal{S}}} ({\bf u}).
\end{align*}
To transition from the third to the fourth line above, we have use the property that $V_{\cal{S}} \subset V_{\cal{A}}$.  $\Box$

\subsubsection{Non-negative Wigner functions} \label{section:hudson_positivity}

To complete the proof of Theorem~\ref{theo:hudson}, we consider a pure state which has non-negative Wigner function and we determine its Wigner function. We will show that this function coincides with the Wigner function of a CSS state. By refining the qudit proof of Gross \cite{Gross} we will show that

\begin{Lemma} \label{prop:positivity_implies_CSS}
If a pure real state $\ket \psi$ has non-negative Wigner function $W_\psi$, then its Wigner function is of the form
\begin{equation}
\label{iso}
W_\psi({\bf u}) = \frac{1}{2^n} \delta_T({\bf u}),
\end{equation}
where $T =  ({\bf p_0} + N^\perp) \times ({\bf q_0} + N)$, ${\bf p_0, q_0}$ are two vectors of $\Z_2^n$ and $N$ is a linear subspace of $\Z_2^n$
\end{Lemma}
The proof of this result comprises the next 5 lemmas. First, we find, by explicit computation that
\begin{Lemma} \label{lemma:hudson_expression_W}
The Wigner function $W_\psi$ of a pure real state $|\psi \rangle$, at some point $(\mathbf{p},\mathbf{q}) \in \Z_2^{2n}$ is
\[ 
W_{\psi}(\mathbf{p},\mathbf{q}) =  \frac{1}{2^n} \sum_{\bf x \in \Z_2^n} (-1)^{\mathbf{(p, x)}}\psi(\mathbf{q})\psi(\mathbf{q}+\mathbf{x}).
\]
where $\psi({\bf x})$ denotes the inner product $\braket{\psi}{\bf x}$.
\end{Lemma}
This result is proved in Appendix~\ref{WiProp}.

This encourages us to study the function $\psi: \Z_2^n \rightarrow \R$ defined by $\psi({\bf x}) = \braket{\psi}{\bf x}$. The support of $\psi$, denoted $\supp(\psi)$, is the set of vectors ${\bf x} \in \Z_2^n$ such that $\psi({\bf x}) \neq 0$.

For fixed ${\bf q}$, we consider the function $K({\bf q}, \cdot)$ defined by
\begin{equation} \label{eqn:K_definition}
K_\psi(\mathbf{q},\mathbf{x}) = \psi(\mathbf{q})\psi(\mathbf{q}+\mathbf{x}).
\end{equation}
It is related to the Wigner function of the state $|\psi\rangle$ via a Fourier transformation
\begin{equation}\label{Fourier}
{\cal F}K_\psi({\bf q}, \cdot) = 2^{n/2} W_\psi(\cdot, {\bf q}).
\end{equation}
The definition of the Fourier transform for the present binary setting is recalled in Appendix~\ref{section:fourier}.

Eq.~(\ref{Fourier}) allows us to relate properties of $|\psi\rangle$ and $W_\psi$.
\begin{Lemma} \label{lemma:hudson_psi_constant}
Let $\ket \psi = \sum_{\bf x} \psi({\bf x}) \ket{\bf x}$ be a pure real state. If $W_\psi$ is non-negative then the function $\psi$ has constant absolute value over its support $\supp(\psi)$.
\end{Lemma}

{\em{Proof of Lemma~\ref{lemma:hudson_psi_constant}.}} By Lemma~\ref{lemma:hudson_expression_W}, $W(\cdot, {\bf q})$ is the Fourier transform of the function $K({\bf q}, \cdot)$ defined in Eq.(\ref{eqn:K_definition}), up to multiplication by $2^{n/2}$. That means that $K({\bf q}, \cdot)$ has non-negative Fourier transform. Therefore, we can apply Bochner's theorem, exactly as stated in Theorem~44 of \cite{Gross} (This result and its proof are unchanged in the binary setting). This proves that the matrix $A^\mathbf{x}_\mathbf{y}=K(\mathbf{q},\mathbf{x}-\mathbf{y})$ is postive semi-definite, where the tuples $\mathbf{x}$ and $\mathbf{y}$ are viewed as the binary writing of the matrix indices. From a well known characterization of postive semi-definite matrices, every principal minor of the matrix $A^\mathbf{x}_\mathbf{y}$ is non-negative. In particular, the determinant
\begin{align*}
\left| 
\begin{array}{ccc}
A^\mathbf{0}_\mathbf{0}&A^\mathbf{0}_\mathbf{x} \\
A^\mathbf{x}_\mathbf{0}&A^\mathbf{x}_\mathbf{x}
\end{array} 
\right|
& =
\left| 
\begin{array}{ccc}
\psi(\mathbf{q})^2&\psi(\mathbf{q})\psi(\mathbf{q}+\mathbf{x})\\
\psi(\mathbf{q})\psi(\mathbf{q}+\mathbf{x})& \psi(\mathbf{q})^2 
\end{array} 
\right|
\end{align*}
is non-negative. This implies the following inequality.
\begin{equation*} 
\psi(\mathbf{q})^4 \geq \psi(\mathbf{q})^2 \psi(\mathbf{q}+\mathbf{x})^2.
\end{equation*}
If ${\bf q} \in \supp(\psi)$ and ${\bf x} \in \Z_2^n$, then we obtain
\begin{equation} \label{eqn:hudson_inequality_psi}
|\psi(\mathbf{q})| \ge|\psi(\mathbf{q}+\mathbf{x})|
\end{equation}
since $\psi(\mathbf{q}) \neq 0$.

Now, consider two vectors $\mathbf{q}$ and $\mathbf{q'}$ of $\supp(\psi)$. Applying Eq.(\ref{eqn:hudson_inequality_psi}) to ${\bf q}$ and ${\bf x= q+q'}$ we find
$
|\psi({\bf q})| \geq |\psi({\bf q+q+q'})| = |\psi({\bf q'})|
$
and exchanging the roles of $\bf q$ and $\bf q'$, we obtain the reverse inequality of (\ref{eqn:hudson_inequality_psi}), and thus 
\begin{equation} \label{eqn:hudson_psi_constant}
|\psi({\bf q})| = |\psi({\bf q'})|.
\end{equation}
This proves that $\psi$ has constant absolute value over its support $\supp(\psi)$. $\Box$ \medskip

\begin{Lemma} \label{lemma:hudson_psi_support}
Let $\ket \psi = \sum_{\bf x} \psi({\bf x}) \ket{\bf x}$ be a pure real state. If $W_\psi$ is non-negative then the support of $\psi$ is an affine subspace of $\Z_2^n$, 
$\supp(\psi) = {\bf q_0} + N$.
\end{Lemma}
 
{\em{Proof of Lemma~\ref{lemma:hudson_psi_support}.}} Let $\bf q, q+x$ and $\bf q+y$ be three vectors in $\supp(\psi)$. We have to show that $\bf q+x+y$ is also in $\supp(\psi)$ (In the qudit case \cite{Gross}, this result is deduced from the qudit version of Eq.~(\ref{eqn:hudson_inequality_psi}). This strategy cannot be adapted here since Eq.~(\ref{eqn:hudson_inequality_psi}) only involves two vectors $\bf q$ and $\bf q+x$.). In order to obtain an equation relating more vectors of $\Z_2^n$, we consider the following $3\times 3$ principal minor of the matrix $A_{\bf y}^{\bf x}$, which is also non-negative by Bochner's theorem.
$$
\left|
\begin{array}{ccc}
A^{\bf 0}_{\bf 0} & A^{\bf 0}_{\bf x} & A^{\bf 0}_{\bf y}\\
A^{\bf x}_{\bf 0} & A^{\bf x}_{\bf x} & A^{\bf x}_{\bf y}\\
A^{\bf y}_{\bf 0} & A^{\bf 0}_{\bf y} & A^{\bf y}_{\bf y}
\end{array}
\right|
\geq 0.
$$
The expansion of this determinant leads to the inequality
\begin{align*}
&\psi({\bf q})^3\left( \psi({\bf q})^3 + 2\psi({\bf q+x})\psi({\bf q+y})\psi({\bf q+x+y}) \right)\\
& - \psi({\bf q})^4 \left(\psi({\bf q+x})^2 -  \psi({\bf q+y})^2 - \psi({\bf q+x+y})^2 \right)
\geq 0
\end{align*}
By contradiction, assume that $\psi({\bf q+x+y}) = 0$, then we have
\begin{equation} \label{eqn:hudson_minor_3x3}
\psi({\bf q})^6 - \psi({\bf q})^4\psi({\bf q+x})^2 -  \psi({\bf q})^4\psi({\bf q+y})^2 \geq 0.
\end{equation}
From Lemma~\ref{lemma:hudson_psi_constant}, the three real numbers $\psi({\bf q})$, $\psi({\bf q+x})$ and $\psi({\bf q+y})$ have the same absolute value. Therefore Eq.(\ref{eqn:hudson_minor_3x3}) cannot be satisfied since the three terms of the left hand side are equal and positive. This contradiction implies that $\psi({\bf q+x+y}) \in \supp(\psi)$. Hence $\supp(\psi)$ is an affine space: $\supp(\psi) = {\bf q_0}+N$ where ${\bf q_0} \in \Z_2^n$ and $N$ is a linear subspace of $\Z_2^n$. $\Box$ \medskip

\begin{Lemma} \label{lemma:hudson_W_q}
Let $\ket \psi = \sum_{\bf x} \psi({\bf x}) \ket{\bf x}$ be a pure real state. If $W_\psi$ is non-negative then for every ${\bf q} \in {\bf q_0} + N$, the function $W_\psi(\cdot, {\bf q})$ is
$$
W_\psi(\cdot, {\bf q}) = c \delta_{\bf p_0 + N^\perp},
$$
where $c = c({\bf q}) \in \R$ and ${\bf p_0 = p_0(q)} \in \Z_2^n$ may both depend on $\bf q$.
Moreover, if ${\bf q} \notin {\bf q_0} + N$, then $W_\psi(\cdot, {\bf q}) = 0$.
\end{Lemma}

{\em{Proof of Lemma~\ref{lemma:hudson_W_q}.}} First, we fix a vector ${\bf q} \in \Z_2^n$ and we focus on the support the function $W_\psi(\cdot, {\bf q})$.
From Lemma~\ref{lemma:hudson_expression_W}, this function satisfies
\begin{align*}
W_{\psi}(\mathbf{p},\mathbf{q}) & =  \frac{1}{2^n} \sum_{\bf x \in \Z_2^n} (-1)^{\mathbf{(p, x)}} \psi(\bf q)\psi(\bf q + \bf x).
\end{align*}
Therefore, $W_\psi(\cdot, {\bf q})$ is the zero function when ${\bf q}$ does not belong to the support of $\psi$, which is ${\bf q_0} + N$ from Lemma~\ref{lemma:hudson_psi_support}.

In what follows, the vector ${\bf q}$ is chosen in ${\bf q_0} + N$. In the above expression of $W_\psi$, the term $\psi(\bf q)\psi(\bf q + \bf x)$ can be replaced by $K(\bf q, \bf x)$, defined in Eq.(\ref{eqn:K_definition}).
The support of the function $K$ is $\supp(K) = (\bf {q_0} + N) \times N$ where $\bf {q_0} + N$ is the support of $\psi$.
Then, $K$ can be restricted to its support. This gives
\begin{align*}
W_{\psi}(\mathbf{p},\mathbf{q}) = \frac{1}{2^n} \sum_{\bf x \in N} (-1)^{\mathbf{(p, x)}} K'(\bf q, \bf x)
\end{align*}
where $K'$ is the restriction of $K$ to its support.

Now note that, for every vector $\bf q \in \Z_2^n$, the function $W_\psi(\cdot, \bf q)$ is constant over the cosets of $N^\perp$. Therefore, this function induces a function $W_\psi([\cdot], {\bf q})$ over $\Z_2^n / (N^\perp)$:
\begin{align*}
W_\psi([\cdot], {\bf q}) : \Z_2^n / (N^\perp) & \longrightarrow \R\\
[{\bf p}] = {\bf p} + N^\perp & \longmapsto W_\psi({\bf p}, {\bf q}).
\end{align*}
The space $\Z_2^n / (N^\perp)$ is isomorphic to the linear space $N$. Indeed, the application from $\Z_2^n$ to the dual $N^*$ of $N$, defined by ${\bf x} \mapsto ({\bf x}, \cdot)$ induces an isomorphism between $\Z_2^n / (N^\perp)$ and $N^*$. Thus, $N^*$ is canonically isomorphic to $N$.

Up to this isomorphism $\Z_2^n / (N^\perp) \simeq N$, the functions $K'(\bf q, \cdot)$ and $W_\psi([\cdot], \bf q)$ are both defined over the same space and $W_\psi([\cdot], {\bf q})$ is the Fourier transform of $K'({\bf q}, \cdot)$ up to multiplication by $2^{n/2}$, that is ${\cal F}K'({\bf q}, \cdot) = 2^{n/2} W_\psi([\cdot], {\bf q})$. Applying $\cal F$ to this equality, we obtain
$$
2^{-n/2} K'({\bf q}, \cdot) = {\cal F}W_\psi([\cdot], {\bf q}),
$$
because $\cal F$ is involutive, from Lemma~\ref{lemma:fourier_involutive} in Appendix~\ref{section:fourier}.

The function $K'(\bf q, \cdot)$ has constant absolute value over $N$ by Lemma~\ref{lemma:hudson_psi_constant}, thus we apply the second item of Bochner Theorem (Theorem~44 in \cite{Gross}) to $W_\psi([\cdot], \bf q)$. This tells us that $W_\psi([\cdot], \bf q)$ is orthogonal to its translations, {\em i.e.}
$$
\sum_{[{\bf p}]} W_\psi([{\bf p}], {\bf q}) W_\psi([\bf p] + [\bf t], {\bf q}) = 0,
$$
for every $[{\bf t}] \in \Z_2^n/(N^\perp)$.
A positive function which satisfies this orthogonality condition can be either zero or proportional to an indicator function $\delta_{[\bf{p_0}]}$. But $W_\psi([\cdot], \bf q)$ cannot be zero. Otherwise $K({\bf q}, \cdot)$ is also the zero function by injectivity of the Fourier transform, and this cannot happen when $\bf q$ is chosen in ${\bf q_0} + N$.  $\Box$ \medskip

The next lemma concludes the proof of Lemma~\ref{prop:positivity_implies_CSS}.
\begin{Lemma} \label{lemma:hudson_W_computation}
Let $\ket \psi = \sum_{\bf x} \psi({\bf x}) \ket{\bf x}$ be a pure real state. If $W_\psi$ is non-negative then $W_\psi$ is of the form
$$
W_\psi = \frac{1}{2^n} \delta_{({\bf p_0} + N^\perp) \times ({\bf q_0} + N)},
$$
where ${\bf p_0, q_0} \in \Z_2^n$ and $N$ is a linear subspace of $\Z_2^n$.
\end{Lemma}

{\em{Proof of Lemma~\ref{lemma:hudson_W_computation}.}} From Lemma~\ref{lemma:hudson_W_q}, the global support, $\supp(W_\psi)$ is the disjoint union
\begin{equation} \label{eqn:hudson_support_W}
\supp(W_\psi) = \bigsqcup_{\bf q \in \bf{q_0} + N} \left( \bf{p_0}({\bf q})+N^\perp \right) \times \{\bf q\}.
\end{equation}
Our first goal is to prove that ${\bf p_0}({\bf q})$ does not depend on ${\bf q}$.
To this end, it is natural to separate the variables ${\bf p}$ and ${\bf q}$ in the writing of $W_\psi$ obtained in Lemma~\ref{lemma:hudson_expression_W}. This leads to
\begin{equation} \label{eqn:hudson_W_separate_variables}
W_\psi({\bf p, q}) = (-1)^{({\bf p, q})} \hat \psi({\bf p}) \psi({\bf q}),
\end{equation}
where $\hat \psi$ is the Fourier transform of $\psi$. Thus the support of $W_\psi$ is also
$
\supp(W_\psi) = \supp(\hat \psi) \times \supp(\psi).
$
This can be satisfied if and only if $\bf p_0$ is independent of $\bf q$ in Eq.(\ref{eqn:hudson_support_W}). This proves that the support of $W_\psi$ is the cartesian product 
$$
\supp(W_\psi) = ({\bf p_0} + N^\perp) \times ({\bf q_0} + N).
$$
Now, let us prove that $W_\psi$ has constant absolute value over its support.
Let $({\bf p, q}) \in \supp(W_\psi)$. Combining Lemma~\ref{lemma:hudson_W_q} and Eq.(\ref{eqn:hudson_W_separate_variables}), we find that the modulus of $W_\psi({\bf p, q})$ is
$$
|c({\bf q})| = |\hat \psi({\bf p})|\cdot|\psi({\bf q})|,
$$
where $c({\bf q})$ is the constant introduced in Lemma~\ref{lemma:hudson_W_q}. Recall that $c({\bf q})$ is independent of ${\bf p}$. We proved in Lemma~\ref{lemma:hudson_psi_constant} that $|\psi({\bf q})|$ is constant, therefore $|c({\bf q})|$ is also independent of ${\bf q}$. This proves that $|c|$ is constant over $\supp(\psi)$. By positivity of $W_\psi$, we have $c = |c|$ and
$$
W_\psi = c \delta_{({\bf p_0} + N^\perp) \times ({\bf q_0} + N)},
$$
for some constant $c \in \R$.

To conclude the proof it remains to evaluate the value of $c$.
By the normalisation of Property 6. of the Wigner function, it suffices to compute the cardinality of the support of $W_\psi$. We find $|\supp(W_\psi)| = |N^\perp|\cdot |N| = 2^{n - \dim N} \cdot 2^{\dim N} = 2^n$, which gives $c = 1/2^n$. This concludes the proof. $\Box$

\subsubsection{Proof of Hudson's theorem for rebits} \label{section:hudson_proof}

Lemma~\ref{prop:computation_wigner_CSS} together with Lemma~\ref{prop:positivity_implies_CSS} enable us to prove a rebit version of Hudson's Theorem.

\medskip
{\em{Proof of Theorem~\ref{theo:hudson}.}} Lemma~\ref{prop:computation_wigner_CSS} implies that every CSS states has non-negative Wigner function.

Now, consider a pure real state $\ket \psi$ which admits a non-negative Wigner function. In order to prove that this is a CSS state, it is enough to prove that its Wigner function coincides with the Wigner function of a pure CSS state $\ket \varphi$. Indeed, since the Wigner function is informationally complete (Property 1.), this implies $\ket \psi = \ket \varphi$.
We proved in Lemma~\ref{prop:positivity_implies_CSS} that $W_\psi$ can be written
$$
W_\psi = \frac{1}{2^n} \delta_{({\bf p_0}+ N^\perp) \times ({\bf q_0} + N)}.
$$
Since $({\bf p_0}+ N^\perp) \times ({\bf q_0} + N) = {\bf t} + V_S$, where
${\bf t} = {\bf (p_0, q_0)}$ and $V_S = N^\perp \times N$, this is indeed the Wigner function of a CSS state by Lemma~\ref{prop:computation_wigner_CSS}. $\Box$

\subsection{Covariance of the rebit Wigner function} \label{section:covariance}

Our next goal is to demonstrate that the action of CSS-ness preserving Clifford gates on Wigner functions $W_\rho$ can be understood simply from the action of such gates on the underlying phase space, c.f. Lemma~\ref{CB} below. To prepare for this result, we make two observations.
\begin{Lemma} \label{prop:conjugation_discretization}
Let $g \in G_{CSS}$. Then, there exists a unique pair $(F, {\bf x})$ composed of a vector ${\bf x} \in \Z_2^{2n}$ and a symplectic matrix $F \in \Sp_{2n}(\Z_2)$ such that
\begin{equation}\label{gTg}
g T_{\bf a} g^\dagger = (-1)^{[{\bf x}, {\bf a}]} T_{F{\bf a}}, \quad \forall {\bf a} \in V_{\cal A}.
\end{equation}
\end{Lemma}
The proof of Lemma~\ref{prop:conjugation_discretization} is given in Appendix~\ref{appsection:CSS_Clifford}.

Furthermore, the action of a $g \in G_{CSS}$ on a translation operator $T_{\textbf{a}} \in {\cal{T}}$ by conjugation  induces a morphism from the CSS Clifford group to the affine group $\AGL_{2n}(\Z_2)$. Recall that an affine transformation of $\AGL_{2n}(\Z_2)$ is an application of the form $A(F, {\bf t}): {\bf a} \mapsto F {\bf a} + {\bf t}$, where $F \in \GL_{2n}(\Z_2)$ is a linear application and ${\bf t}$ is a vector of $\Z_2^{2n}$. In the present work $F$ is often symplectic and this affine map is then called an affine symplectic map. The set of affine symplectic transformations of $\Z_2^{2n}$ is a subgroup of the affine group denoted $\ASp_{2n}(\Z_2^{2n})$.

\begin{Lemma}\label{prop:CSS_morphism}
Let $\cal F$ be the application
\begin{align*}
{\cal F}: G_{CSS} & \longrightarrow \ASp_{2n}(\Z_2)\\
g & \longmapsto A(F, {\bf t})
\end{align*}
such that $g T_{\bf a} g^\dagger = (-1)^{[{\bf t}, F{\bf a}]} T_{F{\bf a}}$, for all ${\bf a}$.
Then $\cal F$ is a group morphism.
\end{Lemma}
The proof of Lemma~\ref{prop:CSS_morphism} is given in Appendix~\ref{appsection:CSS_Clifford}.
The application $\cal F$ is well defined by unicity in Lemma~\ref{prop:conjugation_discretization}.
The translation vector ${\bf t}$ and the vector ${\bf x}$ of Lemma~\ref{prop:conjugation_discretization} are related by the equation ${\bf t} = F {\bf x}$. 

We are now ready to state the covariance result.
 
\begin{Lemma}\label{CB}
The $n$-rebit Wigner function $W$ is covariant under $G_{CSS}$, in the sense that for all $\rho$, for all $\textbf{u}\in \mathbb{Z}_2^{2n}$,  and for all $g \in G_{CSS}$ it holds that
\begin{equation}\label{WifuCov}
W_{g^\dagger \rho g}(\textbf{u}) = W_\rho\left({\cal F}(g)({\bf u})\right).
\end{equation}
\end{Lemma}
Applying this result to $g \rho g^\dagger = (g^{-1})^\dagger \rho g^{-1}$, we find
$$
W_{g \rho g^\dagger}(\textbf{u}) = W_\rho\left({\cal F}(g)^{-1}({\bf u})\right) =  W_\rho\left( F^{-1}({\bf u +  t})\right),
$$
where ${\cal F}(g) = A(F, {\bf t})$.

\medskip
{\em{Proof of Lemma~\ref{CB}.}} Let $g \in G_{CSS}$ and let ${\cal F}(g) = A(F, {\bf t})$ be its induced affine symplectic map.
First, consider the image of $A_{\bf u}$ by conjugation by $g$. Using Eq.(\ref{eqn:A_u_second_form}), we obtain
\begin{align*}
g A_{\bf u} g^{\dagger} & =  \frac{1}{2^n} \sum_{{\bf a} \in V_{\cal A}} (-1)^{[{\bf u}, {\bf a}]} g T_{\bf a} g^\dagger\\
& = \frac{1}{2^n} \sum_{{\bf a} \in V_{\cal A}} (-1)^{[{\bf u}, {\bf a}] + [{\bf t}, F{\bf a}]} T_{F{\bf a}}\\
& = \frac{1}{2^n} \sum_{{\bf a} \in V_{\cal A}} (-1)^{[F{\bf u} + {\bf t}, F{\bf a}]} T_{F{\bf a}}\\
& =\frac{1}{2^n} \sum_{{\bf b} \in V_{\cal A}}  (-1)^{[F{\bf u} + {\bf t}, {\bf b}]} T_{{\bf b}}\\
& = A_{{\cal F}(g)({\bf u})}
\end{align*}
where we have used $[{\bf u}, {\bf a}] = [F{\bf u}, F{\bf a}]$ and the fact that $F$ induces a bijection of the set $V_{\cal A}$.
This leads to
\begin{align*}
2^n W_{g^\dagger \rho g}({\bf u}) & = \Tr(A_{\bf u} g^\dagger \rho g)\\
& = \Tr(g A_{\bf u} g^\dagger \rho)\\
& = \Tr( A_{{\cal F}(g)({\bf u})} \rho)\\
& = 2^n W_\rho({\cal F}(g)({\bf u})),
\end{align*}
which proves the covariance. $\Box$
\medskip

For $n\geq 2$, $W$ is not covariant under all real Clifford operations. As an example, consider $n=2$ and $g=H_1$, which is real Clifford but not CSS-ness preserving. $H_1$ converts a Bell state into a 2-qubit graph state. The former has positive and the latter negative Wigner function. Hence, $H_1$ does not transform $W$ covariantly.

\subsection{Efficient simulation of Clifford circuits} \label{section:simulation}

An operational justification for emphasizing positivity of Wigner functions is the following result \cite{Veitch} for qudits: Circuits of Clifford gates and stabilizer measurements acting on an initial state with non-negative Wigner function can be efficiently simulated classically. The discrete Hudson's theorem \cite{Gross} ensures that for pure states, the simulation method based on Wigner functions has the same scope as the Gottesman-Knill theorem. For mixed states it is an extension of that theorem, since not all states with non-negative Wigner function are mixtures of stabilizer states \cite{Gross}. 

Here we prove an analogue of the result \cite{Veitch} for the rebit Wigner function $W$ defined in Eqs.~(\ref{WFa}), (\ref{mwf}).

\begin{Theorem} \label{prop:classical_simulation}
Every circuit consisting of CSS-ness preserving Clifford unitaries and measurements, acting on a product state $\rho = \bigotimes_{i=1}^n\rho_i$ with non-negative Wigner function $W_{\rho}$, can be efficiently classically simulated.
\end{Theorem}

{\em{Proof of Theorem~\ref{prop:classical_simulation}.}} We describe a simulation method based on sampling. For a quantum state $\rho$ represented by a Wigner function $W_\rho$, the probability of an outcome $s$ corresponding to the POVM element $E(s)$ is
$$
P(s) = \sum_{\bf u} W_\rho({\bf u}) W_{E(s)}({\bf u}).
$$
For the allowed observables $O \in {\cal{O}}$, the POVM elements $E(s) = (I +s\,O)/2$ all have positive Wigner function $W_{E(s)}$. Therefore, $P(s)$ can be efficiently estimated if $W_\rho$ is positive (i.e., is a probability distribution), and can be efficiently sampled from. We show by induction that this is indeed the case for all Wigner functions generated by the above circuits.

First, the initial Wigner function for the state $\rho(0)=\rho_1\otimes \rho_2 \otimes ..\otimes \rho_n$, $W_{\rho(0)} = W_{\rho_1}W_{\rho_2}\cdot .. \cdot W_{\rho_n}$, can be efficiently sampled from. It is positive, and the $W_{\rho_i}$ may be sampled from independently, which is efficient.

Now we show that if the Wigner function $W_{\rho(t)}$ after time step $t$ can be efficiently sampled from, then so can the Wigner function $W_{\rho(t+1)}$ after step $t+1$. We distinguish two cases: (a) $\rho(t+1) = g\rho(t)g^\dagger$, with $g \in G_{CSS}$, and (b) $\rho(t+1) \sim \frac{I + s\,O}{2} \rho(t) \frac{I + s\,O}{2}$, with $O \in {\cal{O}}$, $s =\pm 1$. 

{\em{(a) Unitary evolution.}} The Wigner function transforms covariantly under gates $g \in G_{CSS}$, 
$$
W_{\rho(t+1)}(F_g \textbf{u} + \textbf{t}_g) = W_{\rho(t)}(\textbf{u}).
$$ 
Thus, sampling from $W_{\rho(t+1)}$ can be efficiently reduced to sampling from $W_{\rho(t)}$. In particular, gates in $G_{CSS}$ preserve the positivity of the Wigner function.

{\em{(b) Projective measurement.}} We note 
\begin{Lemma} \label{lemma:simulation_measurement}
The Wigner function of the state $\rho'$ of the system after measuring $T_{\bf a} \in \cal O$ with the outcome $s \in \{\pm 1\}$ is
$$
W_{\rho'}({\bf u}) =
\begin{cases}
\frac{1}{2}\left(W_\rho({\bf u}) + W_{\rho}({\bf u+a}) \right) &\text{ if } s \cdot (-1)^{[{\bf u, a}]} = 1\\
0 & \text{ else}
\end{cases}
$$
where $\rho$ is the state before measurement. In particular, measurements of observables in $\cal O$ preserve the positivity of the Wigner function of the system.
\end{Lemma}
$W_{\rho(t+1)}$ is sampled from as follows. Repeat: (1) Call the sampling routine for $W_{\rho(t)}$, which returns a $\textbf{u} \in V$. (2) Report the measurement outcome $s = (-1)^{[\textbf{u},\textbf{a}]}$. (3)  Flip a fair coin, and, depending on the outcome,  report $\textbf{u}$ or $\textbf{u} + \textbf{a}$ as sample from $W_{\rho(t+1)}$.

This concludes the proof of Theorem~\ref{prop:classical_simulation}, subject to the proof of Lemma~\ref{lemma:simulation_measurement}. $\Box$\medskip

{\em{Remark 1:}} The locality of the initial state, $\rho(0) = \rho_1\otimes \rho_2 \otimes .. \otimes \rho_n$ is of no physical significance. It is just one possible way to ensure that the positive $W_{\rho(0)}$ can be efficiently sampled from by a classical algorithm.

{\em{Remark 2:}} The present simulation method is similar to its qudit counterpart \cite{Veitch2}, but  a difference occurs in measurement. Here, mere positivity of the effect $W_{E(s)}$ and positivity of $W_{\rho_{\text{in}}}$ for the input state $\rho_{\text{in}}$ do not imply positivity of the Wigner function $W_{\rho_{\text{out}}}$ for the output state $\rho_{\text{out}}$. Example: The two-rebit state $\rho=(I+X_1Z_2)/4$ has positive Wigner function, and the POVM-element $(I+Z_1X_2)/2$ is also positively represented. However, the state after measurement, a pure stabilizer state with stabilizer group ${\cal{S}} = \langle X_1Z_2,Z_1X_2\rangle$, has {\em{negative}} Wigner function. Note that $Z_1X_2 \not \in {\cal{O}}$. \medskip

{\em{Proof of Lemma~\ref{lemma:simulation_measurement}.}} For all $T_\textbf{a} \in {\cal{A}}$, $T_\textbf{y} \in {\cal{O}}$, it holds that
\begin{equation}
\label{PlusMul}
\text{if } [T_\textbf{a},T_\textbf{y}]=0,\,\text{then } T_\textbf{a} T_\textbf{y} = T_{\textbf{a}+\textbf{y}}.
\end{equation}
This is a consequence of all $T_\textbf{y} \in {\cal{O}}$ being entirely of $X$-type or $Z$-type (by definition of ${\cal{O}}$).

We define the set ${\cal{A}}_\textbf{y}$ as ${\cal{A}}_\textbf{y}=\{T_\textbf{a} \in {\cal{A}}|\, [\textbf{a},\textbf{y}]=0\}$. It has the property that
\begin{equation}
\label{lcsId}
T_\textbf{y} {\cal{A}}_\textbf{y} = {\cal{A}}_\textbf{y}.
\end{equation}
Eq.~(\ref{lcsId}) holds because $[T_\textbf{a},T_\textbf{y}] =0  \Leftrightarrow [T_\textbf{y}T_\textbf{a},T_\textbf{y}] =0$, and Eq.~(\ref{PlusMul}) ($T_\textbf{y}T_\textbf{a} \in {\cal{A}}$, i.e., has the right sign). 

Now, the update $W_\rho \mapsto W_{\rho'}$ under measurement of the observable $T_\textbf{y} \in {\cal{O}}$, with outcome $s =\pm 1$, is
$$
\begin{array}{l}
W_{\rho'}(\textbf{u}) \sim \displaystyle{\frac{1}{2^n} \text{Tr}\left(A_\textbf{u} \frac{I + s\,T_\textbf{y}}{2} \rho \frac{I + s\,T_\textbf{y}}{2}\right)}\vspace{1mm}\\
= \displaystyle{\frac{1}{2^{2n}} \text{Tr}\left(\frac{I + s\, T_\textbf{y}}{2} T_\textbf{u} \left[\sum_{T_\textbf{a} \in {\cal{A}}}T_\textbf{a}\right]  T_\textbf{u}^\dagger \frac{I + s\,T_\textbf{y}}{2} \rho \right)}\vspace{1mm}\\
= \displaystyle{\frac{1}{2^{2n}} \text{Tr}\left(T_\textbf{u} \frac{I + s\,(-1)^{[\textbf{u},\textbf{y}]} T_\textbf{y}}{2}\left[\sum_{T_\textbf{a} \in {\cal{A}}_\textbf{y}}T_\textbf{a}\right]  T_\textbf{u}^\dagger \rho \right)}\vspace{1mm}\\
= \displaystyle{\frac{1}{2^{2n}} \frac{1+s\,(-1)^{[\textbf{u},\textbf{y}]}   }{2} \text{Tr}\left(T_\textbf{u} \left[\sum_{T_\textbf{a} \in {\cal{A}}_\textbf{y}}T_\textbf{a}\right] T_\textbf{u}^\dagger \rho \right)}\vspace{1mm}\\
= \displaystyle{\frac{\delta_{s,(-1)^{[\textbf{u},\textbf{y}]}}}{2^{n+1}} \text{Tr}\left(\left[ A_\textbf{u} + T_\textbf{y} A_\textbf{u} T_\textbf{y}\right] \rho \right)}\vspace{1mm}\\
= \displaystyle{\frac{\delta_{s,(-1)^{[\textbf{u},\textbf{y}]}}}{2} \left(W_\rho(\textbf{u}) + W_\rho(\textbf{u}+\textbf{y}) \right)}.
\end{array}
$$
When transitioning from the third to the fourth line above, we used the property Eq.~(\ref{lcsId}).  $\Box$

\section{Contextuality}
\label{Cont}

\subsection{Scope of hidden variable models for rebit QCSI}

A quantum-mechanical setting comprising quantum states and measurements is said to be contextual if it cannot be described by any non-contextual hidden variable model. For the rebit scheme of quantum computation by state injection considered here, we first need to determine the scope of the phenomenology that any purported non-contextual HVM must reproduce.

The set of quantum states is unrestricted. The candidate HVM must yield the correct measurement statistics for any real quantum state. However, the observables which can be measured in rebit QCSI, and the sets of observables which can be measured jointly, are restricted. To analyze the situation, we first discuss a few examples, and then impose a general criterion. 

First, the set of observables which can be physically measured in rebit QCSI is ${\cal{O}}=\{X(\textbf{a}_X), Z(\textbf{a}_Z)\}$. The candidate HVM therefore needs to correctly reproduce the probabilities of measurement outcomes for all observables $O \in {\cal{O}}$, and furthermore the correct joint outcome probability distributions for any number of commuting observables in ${\cal{O}}$.

But there is more. For example, consider  the two-rebit observable $X_1 Z_2$, which is in the set ${\cal{A}}$ but not in ${\cal{O}}$. The measurement outcome of $X_1Z_2$ can be obtained by measuring the commuting observables $X_1, Z_2 \in {\cal{O}}$, and then post-processing the outcomes. Therefore,  a measurement of $X_1Z_2$ can be reduced to measurements of commuting observables in ${\cal{O}}$. The same holds for all observables in ${\cal{A}}$. We therefore require that any candidate HVM must reproduce the correct measurement statistics for all observables in ${\cal{A}}$.

We now turn to the simultaneous measurement of compatible observables. Continuing with the above example, it is possible to simultaneously measure the pair of observables $\{X_1,X_1Z_2\}$, namely by the same operations that measured $X_1Z_2$ alone. 

Now, is it possible to simultaneously measure the commuting observables $X_1 Z_2$ and $Z_1X_2$? In the setting of rebit QCSI, this is not the case. The measurement of $X_1Z_2$ necessitates the measurement of $X_1$ and $Z_2$ separately. Since these observables do not commute with $Z_1X_2$, a subsequent measurement of $Z_1X_2$ is no longer guaranteed to reveal the original value. Thus, commuting observables in ${\cal{A}}$ need not be simultaneously measurable in the same way as commuting observables in ${\cal{O}}$. 

Based on the phenomenology discussed above, we adopt the following operational criterion to define the scope of hidden variable models:
\begin{Crit} \label{cr1}
Be $M$ a set of commuting observables. Any hidden variable model describing $M$ must correctly predict the joint probability distribution $p_M$ of measurement outcomes, if for all observables $O \in M$ the outcomes can be simultaneously obtained from measurements on a single copy of the given quantum state.
\end{Crit}
We denote by ${\cal{M}}$ the set of measurement settings $M \subset {\cal{A}}$ admitted by Criterion~\ref{cr1}. Given a quantum state $\rho$ and a set $M$ of compatible observables, we denote by $p_{M,\rho}$ the probability distribution for measurement outcomes corresponding to $M$.
\begin{Def}\label{HVM1}
A hidden variable model describing the physical setting $(\rho,{\cal{M}})$ consists of (a) a non-empty set ${\cal{S}}$ of internal states, (b) a probability distribution $q$ over ${\cal{S}}$, and (c) conditional probabilities $p(\textbf{s}_M|\,\textbf{u})$, $\textbf{u} \in {\cal{S}}$, for outcomes $\textbf{s}_M=(s_1,s_2,.., s_{|M|})$ of measurements in $M$, $M \in {\cal{M}}$, such that 
\begin{itemize}
\item[(i)]{For every $\textbf{u} \in {\cal{S}}$, all observables $O \in {\cal{A}}$ have definite values, $\lambda_\textbf{u}(O) = \pm 1$, and for all $M \in {\cal{M}}$
\begin{equation}
\label{HVM_i}
p(\textbf{s}_M|\,\textbf{u}) = \prod_{i|\,O_i \in M} \delta_{s_i,\lambda_\textbf{u}(O_i)} .
\end{equation}}
\item[(ii)]{For all $M \in {\cal{M}}$, all triples of commuting observables $A, B, AB \in \big\langle M\big \rangle$, and all $\textbf{u} \in {\cal{S}}$, the value assignments are consistent,
\begin{equation}
\label{Valco}
\lambda_\textbf{u}(AB) = \lambda_\textbf{u}(A) \lambda_\textbf{u}(B).
\end{equation}}
\item[(iii)]{Given the quantum state $\rho$, the probability distribution $q_\rho$ reproduces all probability distributions of measurement outcomes; i.e. 
\begin{equation}
p_{M,\rho}(\textbf{s}_M) = \sum_{\textbf{u} \in {\cal{S}}}  p(\textbf{s}_M|\,\textbf{u})\, q_\rho(\textbf{u}),
\end{equation}
for all $M \subset {\cal{M}}$, and all values of $\textbf{s}_M$.}
\end{itemize}
\end{Def}
In Sections~\ref{SecNCC} and \ref{SecSCC} below, we derive necessary and sufficient conditions for the existence of a hidden variable model over ${\cal{M}}$, or, the other way around, for contextuality. These conditions are expressed in terms of the rebit Wigner function. 

We conclude this section with a characterization of the sets $M \in {\cal{M}}$ of simultaneously measurable observables in QCSI that are admitted by Criterion~\ref{cr1}. 
\begin{Lemma}\label{Mchar}
Be $M \subset {\cal{A}}$ a set of commuting observables. Then, $M \in {\cal{M}}$ if and only if $T_\textbf{a}T_\textbf{b}=T_{\textbf{a} + \textbf{b}}$,\,$\forall\, T_\textbf{a},T_\textbf{b} \in M$.
\end{Lemma}
{\em{Remark 3:}} What is excluded here is the possibility of  $T_\textbf{a}T_\textbf{b}=-T_{\textbf{a} + \textbf{b}}$.
\medskip

{\em{Proof of Lemma~\ref{Mchar}.}} ``If'': Assume that a set $M \subset {\cal{A}}$ has the property that $T_\textbf{a}T_\textbf{b}=T_{\textbf{a} + \textbf{b}}$ for all $T_\textbf{a}, T_\textbf{b} \in M$. Since $T_{\textbf{a}+\textbf{b}}=(-1)^{\textbf{a}_X\cdot \textbf{b}_Z} T_\textbf{a} T_\textbf{b}$, it follows that $\textbf{a}_X\cdot \textbf{b}_Z = \textbf{a}_Z\cdot \textbf{b}_X = 0$ (mod 2), for all $T_\textbf{a}, T_\textbf{b} \in  M$.

Therefore, for all $T_\textbf{a} \in M$, the operators $X(\textbf{a}_X)$ and $Z(\textbf{a}_Z)$ commute with all of $M$ and among themselves. They thus generate a CSS stabilizer
$$
S = \big\langle X(\textbf{a}_X), Z(\textbf{a}_Z)|\, T_\textbf{a} \in M\big\rangle.
$$
By construction, $M \subset S$. Therefore, the measurement outcomes for all observables $O \in M$ can be obtained by measuring the set of observables $\{X(\textbf{a}_X), Z(\textbf{a}_Z)|\, T_\textbf{a} \in M\} \subset {\cal{O}}$, and subsequent classical processing. The set $M$ thus satisfies Criterion~\ref{cr1}.

``Only if'': Since physical measurements are restricted to observables on ${\cal{O}}$, the only way of measuring an observable $T_\textbf{a} \in {\cal{A}}$ is to separately measure its $X$-part $X(\textbf{a}_X)$ and $Z$-part $Z(\textbf{a}_Z)$, and then post-process the measurement outcomes. We assume that for a given set $M = \{T_\textbf{a}\} \subset {\cal{A}}$ Criterion 1 holds. Then, $[X(\textbf{a}_X),Z(\textbf{b}_Z)]=0$, for all $T_\textbf{a}, T_\textbf{b} \in M$, or, equivalently, $\textbf{a}_X\cdot \textbf{b}_Z=0$, for all $T_\textbf{a},T_\textbf{b} \in M$. Since $T_{\textbf{a}+\textbf{b}} = (-1)^{\textbf{a}_X\cdot \textbf{b}_Z}T_\textbf{a}T_\textbf{b}$, it follows that $T_{\textbf{a}+\textbf{b}} = T_\textbf{a}T_\textbf{b}$ for all $T_\textbf{a}, T_\textbf{b} \in M$. $\Box$\medskip

For an illustration of Lemma~\ref{Mchar}, we previously argued that $X_1Z_1$ and $Z_1X_2$ cannot be simultaneously measured in rebit QCSI; $\{X_1Z_1,Z_1X_2\} \not\in {\cal{M}}$. Lemma~\ref{Mchar} detects this as follows: If $T_\textbf{a}=X_1Z_2$ and $T_\textbf{b} =Z_1X_2$ then $T_{\textbf{a}+\textbf{b}}=-Y_1Y_2$, and therefore $T_{\textbf{a}+\textbf{b}} = - T_\textbf{a}T_\textbf{b}$.

\subsection{A necessary condition for contextuality}
\label{SecNCC}

\begin{Theorem}
\label{NCC}
The setting $(\rho,{\cal{M}})$ is contextual only if $W_\rho < 0$.
\end{Theorem}

{\em{Proof of Theorem~\ref{NCC}}}. If $W_\rho>0$ then $W_\rho$ {\em{is}} a valid non-contextual HVM for the setting 
$(\rho, {\cal{M}})$. To verify this claim, we need to check that if $W_\rho>0$ then $W_\rho$ provides the constructs (a) - (c) required in Definition~\ref{HVM1}, and that the conditions (i) - (iii) therein are satisfied. 

A projective measurement of a set $M \in {\cal{M}}$ of commuting observables is represented by POVM elements $E({\textbf{s}_M})$,
\begin{equation}
\label{POVMels}
E(\textbf{s}_M) = \prod_{i| T_{\textbf{a}(i)} \in M} \frac{I + s_i T_{\textbf{a}(i)}}{2},
\end{equation}
and $s_i = \pm 1$, for all $i$.
With Eq.~(\ref{Trip}), the probability of obtaining the outcomes $\textbf{s}_M$ in the measurement of the set of observables $M$ is 
$$
p_{M,\rho}(\textbf{s}_M) =\text{Tr}(E(\textbf{s}_M) \rho) = 2^n \sum_{\textbf{u} \in V} W_{E(\textbf{s}_M)}(\textbf{u}) W_\rho(\textbf{u}).
$$
We thus identify (a) $V={\cal{S}}$, (b) $W_\rho = q$, and (c) $2^n W_{E(\textbf{s}_M)}(\textbf{u}) = p(\textbf{s}_M|\textbf{u})$, for all $\textbf{u}$. $V = \mathbb{Z}_2^{2n}$ is a valid state space and $W_\rho$ a valid probability distribution, since by assumption $W_\rho > 0$. 

It remains to show that $W_{E(\textbf{s}_M)} >0$ for all $M \in {\cal{M}}$. First, we compute $W_{E(s)}$ for $E(s) = \frac{I+sT_\textbf{a}}{2}$ and $T_\textbf{a} \in {\cal{A}}$. Using the orthogonality relation $\text{Tr} (T_\textbf{a} T_\textbf{b}) = 2^n \delta_{\textbf{a},\textbf{b}}$, we find that $2^n W_{E(s)}(\textbf{u}) =\delta_{s,(-1)^{[\textbf{u},\textbf{a}]}}$.  Thus, for all observables $T_\textbf{a} \in {\cal{A}}$ and all states  $\textbf{u} \in V$, we obtain the value assignment 
\begin{equation}
\label{ValAssign}
\lambda_\textbf{u}(T_{\textbf{a}}) =(-1)^{[\textbf{u},\textbf{a}]}.
\end{equation} 
We now generalize the above computation of the Wigner function of effects from the observables in ${\cal{A}}$ to all sets $M \in {\cal{M}}$ of measurements. To this end, we note that by Lemma~\ref{Mchar} the POVM elements $E(\textbf{s}_M)$ of Eq.~(\ref{POVMels}) can be rewritten as
$$
E(\textbf{s}_M) = \frac{1}{2^{|M|}} \left( \sum_{N \subset M} \left[\prod_{T_{\textbf{a}(i)} \in N} \!\! s_i \right]  T_{\sum_{T_{\textbf{a}(i)} \in N} \textbf{a}(i)}\right).
$$
Hence we obtain
\begin{equation}
\label{Wels}
2^n W_{E(\textbf{s}_M)}(\textbf{u}) = \prod_{i| T_{\textbf{a}(i)} \in M} \delta_{s_i,(-1)^{[\textbf{u},\textbf{a}(i)]}}.
\end{equation}
Thus, $2^n W_{E(\textbf{s}_M)}$ does indeed represent conditional probabilities, as required for $2^n W_{E(\textbf{s}_M)}(\textbf{u}) = p(\textbf{s}_M|\textbf{u})$.

Regarding (i), the assignment of Eq.~(\ref{ValAssign}) demonstrates that for all states $\textbf{u} \in {\cal{S}}$, all observables in ${\cal{A}}$ have definite values, as required. Furthermore, for this value assignment, the expression Eqs.~(\ref{Wels}) for the conditional probability $p(\textbf{s}_M|\textbf{u})$ matches the required expression Eq.~(\ref{HVM_i}).  

Regarding (ii), the value assignment Eq.~(\ref{ValAssign}) leads to the constraints
$$
\lambda_\textbf{u}(T_{\textbf{a}+\textbf{b}}) = \lambda_\textbf{u}(T_\textbf{a}) \lambda_\textbf{u}(T_\textbf{b}),\;\forall \textbf{u} \in {\cal{S}}, \forall \, T_\textbf{a},T_\textbf{b}, T_{\textbf{a}+\textbf{b}} \in {\cal{A}}.
$$
Since, by Lemma~\ref{Mchar}, $T_{\textbf{a}+\textbf{b}}= T_\textbf{a}T_\textbf{b}$ for all $T_\textbf{a},T_\textbf{b},T_{\textbf{a}+\textbf{b}} \in  \langle M\rangle$, the value assignments of Eq.~(\ref{ValAssign}) are consistent for all $M \in {\cal{M}}$. 

Finally, condition (iii) is satisfied by construction of the Wigner function. 

We have thus shown that if $W_\rho>0$ then $W_\rho$ provides a non-contextual HVM for the setting $(\rho,{\cal{M}})$. The claim follows by negation of this statement. $\Box$\medskip

Finally, as an application of Theorem~\ref{NCC}, we briefly discuss the state-dependent version of Mermin's star \cite{Merm}. Employing a Greenberger-Horne-Zeilinger (GHZ)-state $(|000\rangle + |111\rangle)/\sqrt{2}$ in the rebit setting, there is neither negativity nor contextuality. The GHZ state, being of CSS type, has a non-negative Wigner function and hence, by Theorem~\ref{NCC}, is non-contextual. Correspondingly, Mermin's parity proof does not apply to rebits because the local Pauli observables $Y_i$ are imaginary.

\subsection{A sufficient condition for contextuality}
\label{SecSCC}

Below we provide a sufficient criterion for contextuality in terms of the Wigner function. It involves the notion of an isotropic subspace. A subspace $U \subset V=\mathbb{Z}_2^{2n}$ is {\em{isotropic}} if, for all $\textbf{v},\textbf{w} \in U$, $
[\textbf{v},\textbf{w}]:= (\textbf{v}_X,\textbf{w}_Z)+ (\textbf{v}_Z,\textbf{w}_X) \mod 2=0$. Such a space $U$ is said to be {\em{maximally isotropic}} if it is a maximal isotropic subspace  of $\mathbb{Z}_2^{2n}$ with respect to inclusion.
This happens if and only if the dimension of the isotropic subspace $U$ is $n$.

\begin{Theorem}\label{SCC}
The $n$-rebit setting $(\rho,{\cal{M}})$ is contextual if there exists a maximal isotropic subspace $U\subset \mathbb{Z}_2^{2n}$ and a vector $\nu \in \mathbb{Z}_2^{2n}$ such that
$$
\sum_{\textbf{v} \in U} W_\rho(\textbf{v}+\nu) <0.
$$
\end{Theorem}
Comparing Theorems~\ref{NCC} and \ref{SCC}, we find that our necessary and sufficient conditions for contextuality do not match. This indicates the possibility of a Wigner-negative non-contextual phase. Such a phase does indeed exist, as we show in Section~\ref{NegNeqCon}.\medskip

To prove Theorem~\ref{SCC}, we construct a family of witness functions ${\cal{W}}$ which can detect contextuality. Each such function is based on an isotropic subspace $U \subset\mathbb{Z}_2^{2n}$ with a basis ${\cal{B}}(U) = \{\textbf{a}(1),\textbf{a}(2),.., \textbf{a}(m)\}$, and can be evaluated on points $\textbf{x} \in \mathbb{Z}_2^m$, for any density operator $\rho$. Namely, we define
\begin{equation}\label{cWit}
{\cal{W}}^{{\cal{B}}(U)}_\rho (\textbf{x}) = \left\langle \sum_{\textbf{z} \in \mathbb{Z}_2^m}  \left[ \prod_{i=1}^m (-1)^{z_i x_i} \right] T_{\sum_i z_i \textbf{a}(i)} \right\rangle_\rho.
\end{equation}
The contextuality witnesses ${\cal{W}}^{{\cal{B}}(U)}$ resemble the CSW-witnesses \cite{CSW} in that they are linear operators for which the range of expectation values allowed by quantum mechanics is strictly greater than that allowed for non-contextual HVMs. We make the following observation.

\begin{Lemma}\label{Wit}
The setting $(\rho, {\cal{M}})$ is contextual if there exists an isotropic subspace $U \subset \mathbb{Z}_2^{2n}$  such that ${\cal{W}}^{{\cal{B}}(U)}_\rho < 0$.
\end{Lemma}
Before turning to the proof of Lemma~\ref{Wit}, we illustrate the contextuality witnesses Eq.~(\ref{cWit}) in a specific case.\smallskip

{\em{Example.}} Consider two rebits, and a maximal isotropic subspace $U=\mathbb{Z}_2^2=\text{span}(\{ \textbf{a},\textbf{b}\})$ such that $T_\textbf{a} =X_1Z_2$ and $T_\textbf{b}=Z_1X_2$.  With these specifications,
$$
{\cal{W}}_\rho^{\{\textbf{a},\textbf{b}\}}(\textbf{0}) = \langle I_{12} + X_1Z_2 + Z_1X_2 - Y_1Y_2\rangle_\rho.
$$
Note that $T_{\textbf{a}+\textbf{b}} = -Y_1Y_2= -T_\textbf{a}T_\textbf{b}$. If we choose $\rho=|K_2\rangle\langle K_2|$ for a graph state $|K_2\rangle$ with stabilizer relations $X_1Z_2 |K_2\rangle  = Z_1X_2|K_2\rangle = -|K_2\rangle$, then ${\cal{W}}_{|K_2\rangle}^{\{\textbf{a},\textbf{b}\}}(\textbf{0})=-2$. The witness ${\cal{W}}$ can thus indeed take negative values, but what does that say about contextuality?

To answer this question, assume there exists a non-contextual HVM in which all observables in ${\cal{A}}$ have values $\lambda(\cdot)=\pm 1$,  and that these values satisfy the compatibility condition Eq.~(\ref{Valco}). Then, $\lambda(X_1Z_2)=\lambda(X_1)\lambda(Z_2)$ and  $\lambda(Z_1X_2)=\lambda(Z_1)\lambda(X_2)$. Similarly, $\lambda(-Y_1Y_2)=\lambda(X_1X_2)\lambda(Z_1Z_2)=\lambda(X_1)\lambda(X_2)\lambda(Z_1)\lambda(Z_2)$. Therefore, the HVM-version of the witness ${\cal{W}}{\{\textbf{a},\textbf{b}\}}(\textbf{0})$ evaluates to
$$
{\cal{W}}_{\lambda}^{\{\textbf{a},\textbf{b}\}}(\textbf{0}) = \left((1+\lambda(X_1)\lambda(Z_2)\right)  \left((1+\lambda(Z_1)\lambda(X_2)\right),
$$
and is thus non-negative for every value assignment $\lambda$ to the observables $X_1$, $X_2$, $Z_1$, $Z_2$. Hence, it is also non-negative for all probabilistic mixtures over such assignments. A negative value of ${\cal{W}}_\rho^{\{\textbf{a},\textbf{b}\}}(\textbf{0})$ is therefore an indicator of contextuality.

In addition, we observe that the witness ${\cal{W}}_\rho^{\{\textbf{a},\textbf{b}\}}(\textbf{0})$ is closely related to state-independent contextuality. 
Combining the aforementioned relations for $\lambda(T_\textbf{a}=X_1Z_2)$, $\lambda(T_\textbf{b}=Z_1X_2)$ and $\lambda(T_{\textbf{a}+\textbf{b}}=-Y_1Y_2)$, we find that $\lambda(T_{\textbf{a}+\textbf{b}})=\lambda(T_\textbf{a})\lambda(T_\textbf{b})$. By condition Eq.~(\ref{Valco}), this contradicts with the above operator relation $T_{\textbf{a}+\textbf{b}} = -T_\textbf{a}T_\textbf{b}$, giving rise to a state-independent parity proof of contextuality. In fact, the proof in question is a locally rotated version of Mermin's square \cite{Merm} (also see Eq.~(\ref{RotM})).\medskip

{\em{Proof of Lemma~\ref{Wit}.}} We prove the converse statement, namely that if $(\rho,{\cal{M}})$ is non-contextual then ${\cal{W}}^{{\cal{B}}(U)}_\rho > 0$ for all isotropic subspaces $U \in \mathbb{Z}_2^{2n}$ and all bases thereof.

Assume there exists a non-contextual HVM describing the setting $(\rho,{\cal{M}})$. Then, by property (i) of Definition~\ref{HVM1}, the states of this HVM must have definite values $\pm 1$ for all observables in ${\cal{A}}$. Furthermore, for any state $\textbf{u}$ of the HVM, these values must satisfy the consistency condition (ii) of Definition~\ref{HVM1}. 

Specifically, the set $M=\{Z(\textbf{a}_Z)|\, a_\textbf{Z} \in \mathbb{Z}_2^n\}$ satisfies Criterion~\ref{cr1}. Therefore by Property (ii) of Def.~\ref{HVM1}, $\lambda_\textbf{u}\left(T_{(\textbf{a}_Z,0)}\right) = \lambda_\textbf{u}(Z(\textbf{a}_Z)) = \prod_{i|[\textbf{a}_Z]_i=1}\lambda_\textbf{u}(Z_i)$.
Likewise,  $\lambda_\textbf{u}\left(T_{(0,\textbf{a}_X)}\right) = \lambda_\textbf{u}(X(\textbf{a}_X)) = \prod_{i|[\textbf{a}_X]_i=1}\lambda_\textbf{u}(X_i)$. Analogously, for any $T_{(\textbf{a}_Z,\textbf{a}_X)} \in {\cal{A}}$, the set $M=\{T_{(\textbf{a}_Z,0)}, T_{(0,\textbf{a}_X)}, T_{(\textbf{a}_Z,\textbf{a}_X)}\}$ satisfies Criterion~\ref{cr1},  since by definition of ${\cal{A}}$ the Pauli operators $T_{(\textbf{a}_Z,0)}$,  $T_{(0,\textbf{a}_X)}$ commute, and $T_{(\textbf{a}_Z,0)} T_{(0,\textbf{a}_X)}=T_{(\textbf{a}_Z,\textbf{a}_X)}$. Therefore, by Eq.~(\ref{Valco}), 
$\lambda_\textbf{u}\left(T_{(\textbf{a}_z,\textbf{a}_x)}\right) = \lambda_\textbf{u}\left(T_{(\textbf{a}_z,0)}\right)\lambda_\textbf{u}\left(T_{(0,\textbf{a}_x)}\right)$. 

Combining the above three relations, we find that for all $T_\textbf{a} \in {\cal{A}}$, the value $\lambda(T_\textbf{a})$ follows from the values $\lambda(X_i)$, $\lambda(Z_i)$  assigned to the local observables $X_i$ and $Z_i$, for $i=1,..,n$. We may write this as
\begin{equation}
\label{ValAssignGen}
\lambda_\textbf{u}(T_{\textbf{a}}) =(-1)^{[\textbf{u},\textbf{a}]}, \; \forall \textbf{u} \in {\cal{S}},
\end{equation} 
and ${\cal{S}}=\mathbb{Z}_2^{2n}$. We find that the same relation Eq.~(\ref{ValAssign}) which held for HVMs derived from the Wigner function holds for all non-contextual HVMs.

As a consequence, for all $\textbf{u} \in {\cal{S}}$, it holds that
$$
\lambda_\textbf{u}\left(T_{\textbf{a}+\textbf{b}}\right) = \lambda_\textbf{u}\left(T_{\textbf{a}}\right)\lambda_\textbf{u}\left(T_{\textbf{b}}\right),\, \forall\, T_\textbf{a},T_\textbf{b}, T_{\textbf{a} + \textbf{b}} \in {\cal{A}}.
$$
We rewrite this condition as
\begin{equation}
\label{coa}
\lambda_\textbf{u}\!\left(T_{\textbf{a}+\textbf{b}}\right) = \lambda_\textbf{u}\!\left(T_{\textbf{a}}\right)\lambda_\textbf{u}\!\left(T_{\textbf{b}}\right)\!,\, \forall\, T_\textbf{a},T_\textbf{b}\!\! \in\! {\cal{A}}\,\,\text{s.th.}\,\, [T_\textbf{a},T_\textbf{b}]\!=\!0.
\end{equation}
We now evaluate the witness ${\cal{W}}^{{\cal{B}}(U)}_\rho(\textbf{x})$ under the assumption of a non-contextual HVM. Assuming the system is in the state $\textbf{u} \in {\cal{S}}$ of the HVM, and using the property Eq.~(\ref{coa}), the witness of Eq.~(\ref{cWit}) becomes
$$
\begin{array}{rcl}
{\cal{W}}^{{\cal{B}}(U)}_{\lambda_\textbf{u}}(\textbf{x}) &=& \displaystyle{\sum_{\textbf{z} \in \mathbb{Z}_2^m}  \left[ \prod_{i=1}^m (-1)^{z_i x_i} \right] \lambda_\textbf{u}(T_{\sum_i z_i \textbf{a}(i)})}\\
&=& \displaystyle{\sum_{\textbf{z} \in \mathbb{Z}_2^m}   \prod_{i=1}^m \left( (-1)^{x_i} \lambda_\textbf{u}(T_{\textbf{a}(i)})\right)^{z_i} }\\
&=& \displaystyle{\prod_{i=1}^m \left(1+  (-1)^{x_i} \lambda_\textbf{u}(T_{\textbf{a}(i)}) \right)}  \vspace{1mm}\\
&\geq & \displaystyle{0}.
\end{array}
$$
In transitioning from the first to the second line above, we have used the property that $U=\text{span}(\{\textbf{a}(i)\})$ is isotropic, such that Eq.~(\ref{coa}) can be applied.

As a result of the above inequality, for any probability distribution $q_\rho$ over ${\cal{S}}$, the prediction of any non-contextual HVM is
$$
{\cal{W}}^{{\cal{B}}(U)} \geq 0,
$$
for all isotropic subspaces $U \subset \mathbb{Z}_2^{2n}$. The negation of this statement proves the claim. $\Box$\medskip

\noindent
{\em{Remark 4:}} The connection between the witnesses ${\cal{W}}^{{\cal{B}}(U)}$ and state-independent contextuality observed in the earlier two-rebit example persists in the general case.  While the witnesses measure---as it is their purpose---contextuality possessed by quantum states, they are linked to state-independent parity proofs of contextuality as given in \cite{Merm}. Namely, a witness ${\cal{W}}^{{\cal{B}}(U)}$ can assume a negative value only if the associated isotropic space $U$ contains two vectors $\textbf{a}$, $\textbf{b}$ such that $T_{\textbf{a}+\textbf{b}}=-T_\textbf{a} T_\textbf{b}$. Whenever that happens, a parity proof can be built from $T_{\textbf{a}+\textbf{b}}$, $T_\textbf{a}$, $T_\textbf{b}$ and Pauli operators $X_i$, $Z_i$; c.f. Eq.~(\ref{coa}).\medskip

We now relate the witnesses ${\cal{W}}$ to the rebit Wigner function.
\begin{Lemma}\label{WitWifu}
Consider an isotropic subspace $U \subset \mathbb{Z}_2^{2n}$ with basis ${\cal{B}}(U)= \{\textbf{a}(1), \textbf{a}(2),..,\textbf{a}(m)\}$, and a set $\tilde{\cal{B}}=\{\textbf{b}(1), \textbf{b}(2),..,\textbf{b}(m)\}$ such that $[\textbf{a}(i),\textbf{b}(j)] =\delta_{ij}$ for all $i,j=1,..,m$. For every $\eta(\textbf{x}) = \sum_i x_i \textbf{a}(i) \in U$, denote by $\overline{\eta}(\textbf{x})$ the vector $\overline{\eta}(\textbf{x})= \sum_i x_i \textbf{b}(i) \in \mathbb{Z}_2^{2n}$. Then,
$$
{\cal{W}}^{{\cal{B}}(U)}_\rho(\eta(\textbf{x})) = 2^m \sum_{\textbf{v} \in U^\perp} W_\rho(\textbf{v} + \overline{\eta}(\textbf{x})).
$$
\end{Lemma}

{\em{Proof of Lemma~\ref{WitWifu}.}} We may rewrite the witness function ${\cal{W}}$ defined in Eq.~(\ref{cWit}) in terms of $\eta$, $\overline{\eta}$ as 
\begin{equation}\label{Wit2}
{\cal{W}}_\rho^{{\cal{B}}(U)} (\eta) = \left\langle T_{\overline{\eta}} \left(\sum_{\textbf{u} \in U} T_\textbf{u}\right)  T_{\overline{\eta}}^\dagger \right \rangle_\rho.
\end{equation}
We may further rewrite this expression as
$$
\begin{array}{rcl}
{\cal{W}}_\rho^{{\cal{B}}(U)} (\eta) &=& \displaystyle{\frac{2^m}{2^{2n}}\left\langle T_{\overline{\eta}} \left[ \sum_{\textbf{v} \in U^\perp} T_\textbf{v} \left(\sum_{\textbf{u} \in \mathbb{Z}_2^{2n}} T_\textbf{u}\right)  T_\textbf{v}^\dagger \right] T_{\overline{\eta}}^\dagger \right \rangle_\rho}\\
&=& \displaystyle{\frac{2^m}{2^{n}}\left\langle T_{\overline{\eta}} \left[ \sum_{\textbf{v} \in U^\perp} T_\textbf{v} A_0  T_\textbf{v}^\dagger \right] T_{\overline{\eta}}^\dagger \right \rangle_\rho}\\
&=& \displaystyle{2^m  \sum_{\textbf{v} \in U^\perp}  W_\rho(\textbf{v} + \overline{\eta}),}
\end{array}
$$
which demonstrates the claimed relation. In transitioning from the second to the third line above, we have used the fact that $\rho$ is real, and thus $\text{Tr} \,T_\textbf{a}\rho =0$, for all $\textbf{a}$ with $(\textbf{a}_X,\textbf{a}_Z) \mod 2 =1$. $\Box$\medskip

{\em{Proof of Theorem~\ref{SCC}.}} The combined conclusion of Lemmas~\ref{Wit} and \ref{WitWifu} is  that the $n$-rebit setting $(\rho,{\cal{M}})$ is contextual if there exists an isotropic subspace $U\subset \mathbb{Z}_2^{2n}$ with orthogonal complement $U^\perp$ and a vector $\nu \in \mathbb{Z}_2^{2n}$ such that
\begin{equation}
\label{scc1}
\sum_{\textbf{v} \in U^\perp} W_\rho(\textbf{v}+\nu) <0.
\end{equation}
We can further simplify this condition. Suppose that $\sum_{\textbf{v} \in U^\perp} W_\rho(\textbf{v}+\nu) \geq 0$ holds for all $\nu \in \mathbb{Z}_2^{2n}$ when $U$ is maximally isotropic in $\mathbb{Z}_2^{2n}$. Then the same holds for all isotropic subspaces of $\mathbb{Z}_2^{2n}$. To verify this claim, consider a maximally isotropic space $U$ and an isotropic subspace $\tilde{U}$ of $U$. Then, there exists a space $\overline{U} \subset \mathbb{Z}_2^{2n}$ such that $\tilde{U}^\perp = U^\perp \oplus \overline{U}$. Hence,
$$
\sum_{\textbf{v} \in \tilde{U}^\perp} W_\rho(\textbf{v}+\nu) = \sum_{\textbf{v}' \in \overline{U}} \left(\sum_{\textbf{v} \in U^\perp} W_\rho(\textbf{v}'+ \textbf{v}+\nu)\right) .
$$
If every term in brackets on the rhs is $\geq 0$, so is the lhs. Since every isotropic $\tilde{U}$ can be embedded in a maximally isotropic $U$, the above claim follows. That is, we may restrict the condition Eq.~(\ref{scc1}) to maximally isotropic subspaces $U$. In those cases, $U^\perp = U$, which yields the condition stated in Theorem~\ref{SCC}. $\Box$\medskip

\begin{figure}
\begin{center}
\includegraphics[width=7cm]{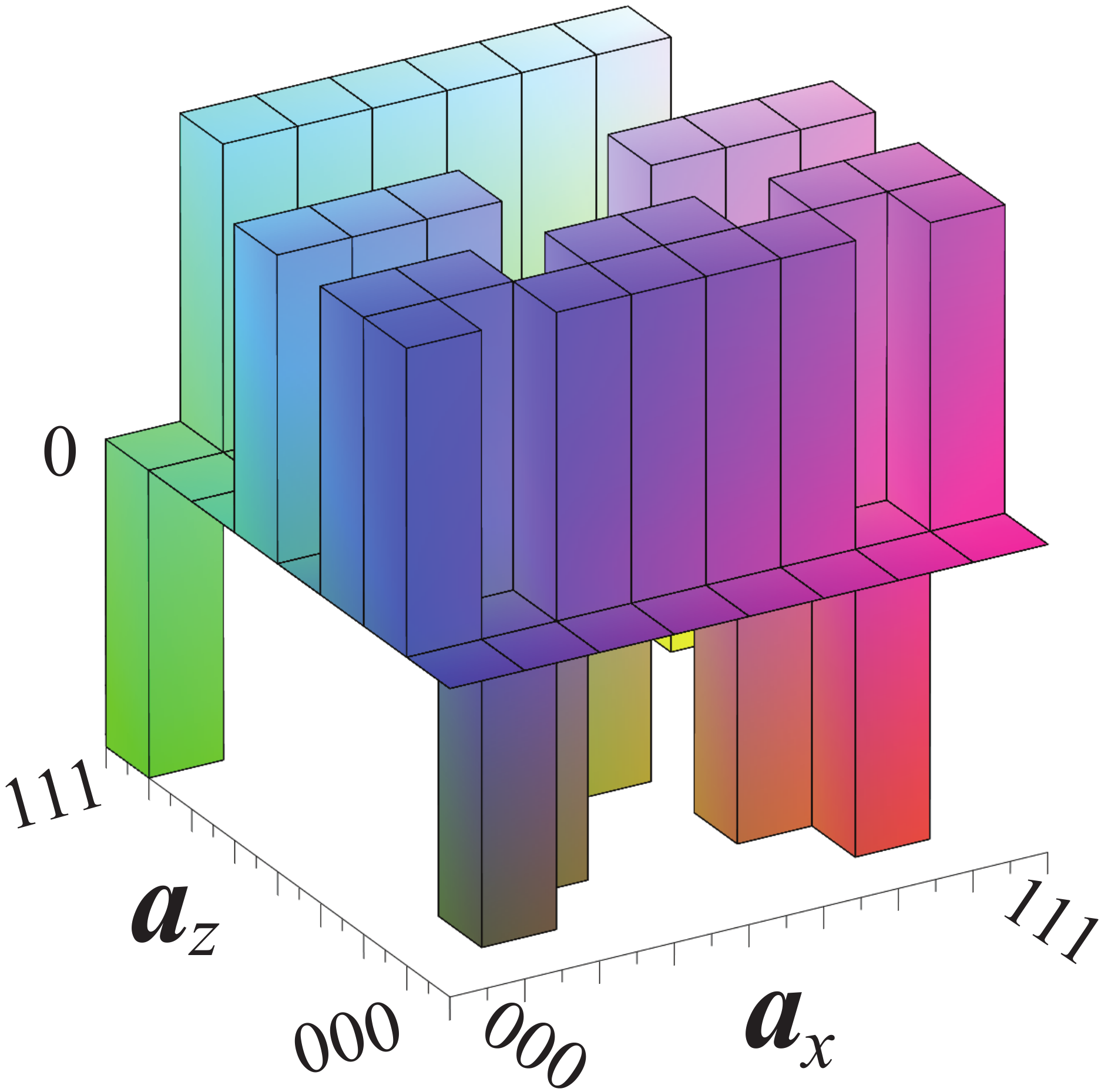}
\end{center}
\caption{\label{K3}Wigner function of the three-rebit graph state $|K_3\rangle$ corresponding to the complete graph $K_3$. The Wigner function takes negative values, and furthermore this negativity is strong enough to witness contextuality of $|K_3\rangle$ with respect to CSS-ness preserving Pauli measurements. }
\end{figure}

Finally, as an application of Theorem~\ref{SCC}, we briefly discuss the state-dependent version of Mermin's star \cite{Merm}, in a locally rotated form. It comprises the nonlocal observables $XXX$, $XZZ$, $ZXZ$, $ZZX$ and local observables $X_i$, $Z_i$, for $i=1..3$. Further, the rotated GHZ-state is a 3-rebit graph state $|K_3\rangle$, with $K_3$ being the fully connected graph of three vertices; hence $|K_3\rangle$ is a joint eigenstate of the above four non-local observables. $W_{|K_3\rangle}$ takes negative values; See Fig.~\ref{K3}. $W_{|K_3\rangle}$ is in fact so negative that it implies contextuality of $|K_3\rangle$ by Theorem~\ref{SCC}. To see this, for the maximal isotropic subspace $U$ appearing in the condition of Theorem~\ref{SCC}, use $U=\text{span}(\{\textbf{a},\textbf{b},\textbf{c}\})$ with $T_\textbf{a}=XZZ$, $T_\textbf{b}=ZXZ$ and $T_\textbf{c}=ZZX$. Correspondingly, in contrast to the original version discussed in Section~\ref{SecNCC}, the rotated version of Mermin's star fully embeds into real quantum mechanics, such that Mermin's parity proof of contextuality applies there. The state-dependent version of this proof applies to rebit QCSI.

\subsection{Are negativity and contextuality the same?}
\label{NegNeqCon}

We observe that the sufficient condition for contextuality in Theorem~\ref{SCC} does in general not match the necessary condition of Theorem~\ref{NCC}. This means that either the sufficient condition is not optimal, or, for the present setting, contextuality and negativity are inequivalent.

\begin{figure}
\begin{center}
\includegraphics[width=5.8cm]{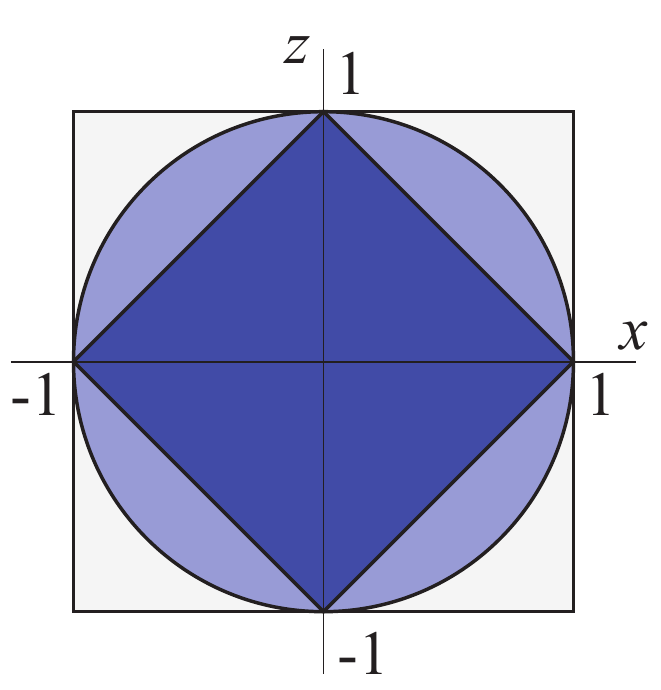}
\caption{\label{shape_B} Phase diagram for the families of states $\tilde{\rho}(x,y)$ of Eq.~(\ref{rho2}), with $x,y \in \mathbb{R}$. Medium shade: the physical states, dark shade: the states classified as non-contextual by Theorem~\ref{NCC}.   The states classified as contextual by Theorem~\ref{SCC} lie outside the square of $|x|,|z|\leq 1$, and are thus not physical.}
\end{center}
\end{figure}

To address the question, we consider the general one-rebit state
\begin{equation}
\label{rho2}
\tilde{\rho}(x,z) = \frac{I + x\, X + z\, Z}{2}.
\end{equation}
The corresponding phase diagram is depicted in Fig.~\ref{shape_B}. The set of physical states is constrained by $
x^2 + z^2 \leq 1$.
By Theorem~\ref{NCC}, $\tilde{\rho}(x,z)$ is non-contextual if $|x|+ |z| \leq 1$, and, 
by Theorem~\ref{SCC}, contextual if  $|x|>1\, \vee\, |z|>1$. We thus find that not a single physical one-rebit state can be  classified as guaranteed contextual by Theorem~\ref{SCC}. 

But this is not a failure of Theorem~\ref{SCC} to get traction. For single qubits, non-contextual HVMs can be constructed \cite{Bell}, \cite{Merm}, and they imply non-contextual HVMs for single rebits as a special case. The states $\tilde{\rho}(x,y)$ with $x^2+z^2\leq1$ and $|x|+|z|>1$ are thus negatively represented but non-contextual. Thus, for the present rebit setting, Wigner function negativity and contextuality are not the same.\medskip

We have to explain how our finding relates to the result by Spekkens \cite{Spekkens} that negativity and contextuality, when suitably defined,  {\em{are}} equivalent notions of non-classicality. In \cite{Spekkens}, the following observations are made: (i) Non-negativity in the quasiprobability distributions representing quantum states is not sufficient for classicality; the conditional probabilities representing measurements must also be non-negative. (ii) A classical explanation cannot be ruled out by considering a single quasiprobability representation; negativity must be demonstrated for all such representations. (iii) The requirement of outcome determinism for sharp measurements should be dropped from the definition of non-contextuality. That is, given an internal state $\textbf{u} \in {\cal{S}}$ of the HVM, the conditional probabilities $p(\textbf{s}_M|\textbf{u})$ for the measurement outcomes are not required to be $\delta$-distributions.

Our setting satisfies the above criterion (i). All projectors onto eigenspaces of the measurable observables $O \in {\cal{O}}$ are non-negatively represented. This is important for the efficient classical simulation method for states with non-negative Wigner function evolving under CSS-ness preserving operations (c.f. Section~\ref{section:simulation}).

Regarding (iii), here we keep the requirement of outcome determinism. Hence, all conditional probability distributions $p(\textbf{s}_M|\textbf{u})$ for measurement outcomes given a fixed internal state are $\delta$-distributions, c.f. Eq.~(\ref{HVM_i}) in Definition~\ref{HVM1}. While not as general as \cite{Spekkens}, it is in accordance with \cite{CSW} (the contextuality measures employed in the qudit counterpart \cite{Howard} of the present work), and \cite{Merm}.

In addition, we point out that not any $\delta$-distribution will do for $p(\textbf{s}_M|\textbf{u})$. Rather, the $\delta$-distributions Eq.~(\ref{HVM_i}) are constrained by outcome compatibility, Eq.~(\ref{Valco}).

Regarding (ii), in contrast to \cite{Spekkens} here we consider only a single quasiprobability distribution---the Wigner function defined in Eq.~(\ref{WFa}). This is motivated by the present computational setting to which the notions of negativity and contextuality are applied: QCSI. As described in Sections~\ref{Reb} and~\ref{CSS}, CSS-states, the observables in ${\cal{O}}$ and the CSS-ness preserving unitaries form a classical reference structure for QCSI on rebits. This implies in particular that, for the present setting, certain bases of Hilbert space are preferred over others for state preparation and measurement. This inequivalence caries over to quasiprobability distributions.

In our setting, a classical explanation {\em{can}} be ruled out by considering a single quasiprobability representation. While mere negativity of the Wigner function is no guarantee for contextuality, a setting $(\rho,{\cal{M}})$ is contextual, hence non-classical, if the Wigner function $W_\rho$ is sufficiently negative to satisfy the condition of Theorem~\ref{SCC}.\medskip

\begin{figure}
\begin{center}
\includegraphics[width=5.8cm]{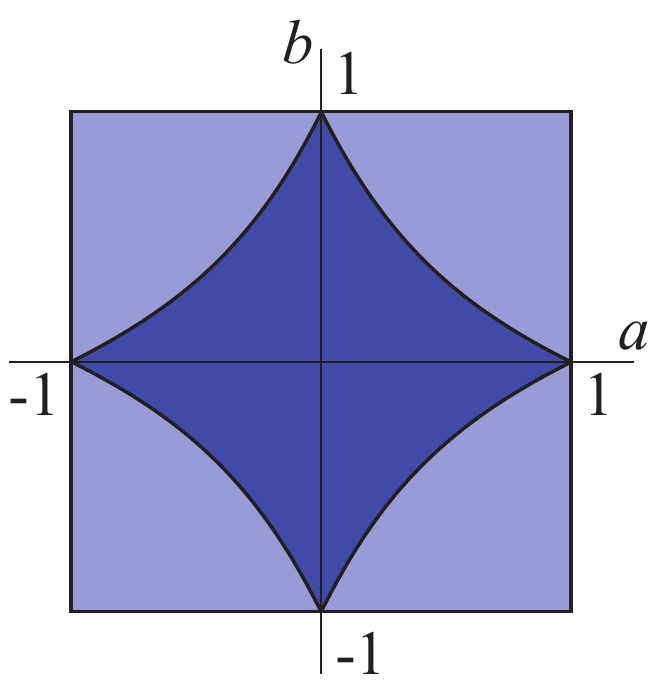}
\caption{\label{shape_A} Phase diagram for the states $\rho(a,b)$ of Eq.~(\ref{rho1}), with $a,b \in \mathbb{R}$. Medium shade: the physical states, dark shade: the states classified as non-contextual by Theorem~\ref{NCC}.}
\end{center}
\end{figure}

Having established that, for the present situation, negativity and contextuality are not equivalent, we turn to the question of whether there are at least large families of states for which the two notions agree.  An example is the family of two-rebit states
\begin{equation}
\label{rho1}
\rho(a,b) = \frac{(I+a\,X_1Z_2)(I+b\,Z_1X_2)}{4}.
\end{equation}
In this case, the conditions of Theorems~\ref{NCC} and \ref{SCC} for contextuality both read
$$
1+\alpha\, a + \beta\, b- \alpha\beta\, ab <0,
$$
for all combinations of $\alpha,\beta =\pm 1$. The corresponding phase diagram is depicted in Fig.~\ref{shape_A}. The physical states fill the square with $|a|,|b|\leq 1$. The corners of that square represent the joint eigenstates of the Pauli operators $XZ$ and $ZX$, and they sit deep in the contextual phase. This fits with our earlier observation that the commuting observables $XZ$ and $ZX$ cannot be simultaneously measured in rebit QCSI. Hence their joint eigenstates cannot be prepared by the restricted gates.\smallskip

This example generalizes as follows.
\begin{Lemma}
\label{TC}
Be $\rho$ a state diagonal in a real stabilizer eigenbasis. Then, $\rho$ is contextual if and only if $W_\rho<0$.
\end{Lemma}

{\em{Proof of Lemma~\ref{TC}.}} Denote by $S_U=\{T_\textbf{v}|\, \textbf{v} \in U\}\subset {\cal{A}}$ the stabilizer in whose joint eigenbasis the state $\rho$ is diagonal; i.e., the corresponding maximal isotropic subspace $U$ is such that $T_\textbf{v} \rho T_\textbf{v}^\dagger = \rho$ for all $\textbf{v} \in U$.

Then, by covariance of the Wigner function under translations, 
$$
W_\rho(\nu) = W_{T_\textbf{v} \rho T_\textbf{v}^\dagger}(\nu) = W_\rho(\textbf{v}+\nu),\;\;\forall \textbf{v} \in U.
$$
In this case, the expression on the lhs of the condition in Theorem~\ref{SCC} simplifies to
$$
\sum_{\textbf{v}\in U} W_\rho(\textbf{v} + \nu) = 2^n W_\rho(\nu).
$$
And thus, Theorem~\ref{SCC} itself simplifies to the statement that if 
$
W_\rho(\nu) <0
$
for some $\nu \in \mathbb{Z}_2^{2n}$ then $\rho$ is contextual. This combined with Theorem~\ref{NCC} proves the claim. $\Box$\medskip

To summarize, unlike for qudits in odd prime dimension \cite{Howard}, for rebits contextuality and Wigner function negativity are not the same. Yet they coincide on all states that are diagonal in a real stabilizer basis. However, note that the definition of contextuality in \cite{Howard} is different from ours. Specifically, in \cite{Howard}, one-qudit states can be classified as contextual based on two-qudit measurements of the given state and a completely depolarized ancilla. 

\section{Contextuality and negativity in quantum computation}
\label{CNC}

\subsection{Resources}

We are now prepared to establish contextuality and Wigner function negativity as necessary resources for universality of QCSI on rebits.
 
 \begin{Theorem}\label{T1}
In quantum computing via state injection on rebits, contextuality of the initial state is necessary for computational universality.
\end{Theorem}
Furthermore,
\begin{Cor}\label{Cor}
In quantum computing via state injection on rebits, Wigner function negativity of the initial state is necessary for computational universality.
\end{Cor}
Corollary~\ref{Cor} is the combination of Theorems~\ref{NCC} and \ref{T1}. 
\medskip

In preparation for the proof of Theorem~\ref{T1}, we note that the witness functions ${\cal{W}}^{{\cal{B}}(U)}$ transform covariantly under CSS-ness preserving unitaries, similar to the Wigner function. Namely, every CSS-ness preserving unitary $g$ can be written as $g = T_\textbf{a} g_F$, where $g_F T_\textbf{b} g_F^\dagger = T_{F\textbf{b}}$ for all $T_\textbf{b} \in {\cal{A}}$, and $[F\textbf{b},F\textbf{c}]=[\textbf{b},\textbf{c}]$ for all $\textbf{b},\textbf{c} \in V$. Then,
using the form Eq.~(\ref{Wit2}) of the contextuality witnesses,
\begin{equation}\label{TransWit}
{\cal{W}}_\rho^{{\cal{B}}(U)}(\eta) = {\cal{W}}^{F^{-1}{\cal{B}}(U)}_{g^{-1}\rho g}(\eta+\overline{\textbf{a}}).
\end{equation}
On the r.h.s., $F^{-1} {\cal{B}}(U)$ is again the basis of an isotropic subspace, since $F$, $F^{-1}$ preserve the commutation relations. In result, for two density matrices $\rho$ and $\rho'$ related by a CSS-ness preserving Clifford unitary, if there is a witness ${\cal{W}}$ that evaluates to $x$ on $\rho$ then there is a witness ${\cal{W}}'$ that evaluates to the same value $x$ on $\rho'$.\medskip

{\em{Proof of Theorem~\ref{T1}.}} If the discussed computational scheme is universal, it must in particular be capable of creating an encoded graph state $|\overline{G}_2\rangle$, with stabilizer $\langle \overline{X}\overline{Z},\overline{Z}\overline{X}\rangle$. Therein, the encoding is that of Rudolph and Grover stated in Eq.~(\ref{rebit_E}),
$$
\sum_k r_k e^{i\phi_k} |k\rangle \longrightarrow \sum_k r_k |k\rangle \otimes \left(\cos \phi_k |0\rangle_A + \sin \phi_k |1\rangle_A \right). 
$$ 
For this encoding, for all qubits $i=1..n$ we have 
\begin{equation}
\overline{X}_i =X_i,\; \overline{Z}_i = Z_i,\; \overline{Y}_i = Y_i \otimes Y_A,
\end{equation}
where $Y:=iXZ$.
With $\overline{i \, I} = i\,Y_A$, this is compatible with the Pauli multiplication table $\overline{Y} = \overline{iZX} = \overline{i\,I}\, \overline{X} \,\overline{Z}$.

All observables in ${\cal{A}}$ have an even number of $Y$'s, and therefore
\begin{equation}
\overline{T} = T,\;\; \forall T \in {\cal{A}}.
\end{equation}
For the state $|\overline{G}_2\rangle$, the contextuality witness based on the operators $T_\textbf{a}=XZ$ and $T_\textbf{b}=ZX$ is negative, namely
\begin{equation}\label{wit}
{\cal{W}}^{\{\textbf{a},\textbf{b}\}}_{|\overline{G}_2\rangle} ((1,1)) = \langle \overline{G}_2|  I -X_1Z_2 - Z_1X_2 -Y_1Y_2 |\overline{G}_2\rangle = -2.
\end{equation}
For two-dimensional isotropic subspaces $U$, $-2$ is the most negative value that a witness ${\cal{W}}^{{\cal{B}}(U)}$ can yield. The final state $|\overline{G}_2\rangle$ thus reveals contextuality maximally.\smallskip

We now prove that also the initial state  fed into the computation must reveal contextuality maximally. The proof is by induction. We consider the circuit which created the state $|\overline{G}_2\rangle$, and assume the gates are performed sequentially, one in each step $m$. We show that if the state $\rho (m)$ after step $m$ reveals contextuality maximally then so does the state $\rho(m-1)$ after step $m-1$. That is, if there exists a witness ${\cal{W}}$ such that ${\cal{W}}_{\rho(m)}(\eta)=-2$ then there exists another witness ${\cal{W}}'$ such that ${\cal{W}}'_{\rho(m-1)}(\eta')=-2$. 

For the gates in the circuit, we distinguish between unitaries and projective measurements. {\em{Case i:}} the gate in step $m$ is a unitary. Then, by construction of the computational scheme, the gate is a CSS-ness preserving Clifford unitary. Then, the claim of the induction step follows from the covariance of the witness functions, Eq.~(\ref{TransWit}).

{\em{Case ii:}} The gate in step $m$ is a projective measurement. Then, by construction of the computational scheme, it is the measurement of an observable $T_\textbf{c} \in {\cal{O}}$. Let the witness for the state $\rho(m)$ be constructed from  the isotropic subspace spanned by $\{\textbf{a}(m),\textbf{b}(m)\}$, such that ${\cal{W}}_{\rho(m)}^{\{\textbf{a}(m),\textbf{b}(m)\}}(\eta)=-2$, for some $\eta$. There are two sub-cases to consider. 

{\em{Case ii/a:}} $T_\textbf{c}$ commutes with both $T_{\textbf{a}(m)}$ and $T_{\textbf{b}(m)}$. Then the value of the witness ${\cal{W}}^{\{\textbf{a}(m),\textbf{b}(m)\}}(\eta)$ is the same for $\rho(m)$ and $\rho(m-1)$, hence $\rho(m-1)$ reveals contextuality maximally.

{\em{Case ii/b:}} $T_\textbf{c}$ does not commute with both $T_{\textbf{a}(m)}$ and $T_{\textbf{b}(m)}$. Then, $T_\textbf{c}$ anti-commutes with two of the three operators $T_{\textbf{a}(m)}$, $T_{\textbf{b}(m)}$, $T_{\textbf{a}(m)+\textbf{b}(m)}$, and commutes with the third. Wlog assume $T_\textbf{c}$ anti-commutes with $T_{\textbf{a}(m)}$ and $T_{\textbf{b}(m)}$, and commutes with $T_{\textbf{a}(m)+\textbf{b}(m)}$. Then, $\left\langle T_{\textbf{a}(m)} \right\rangle_{\rho(m)} = \left\langle T_{\textbf{b}(m)} \right\rangle_{\rho(m)} = 0$. The witness for the state $\rho(m)$ therefore reduces to ${\cal{W}}_{\rho(m)}^{\{\textbf{a}(m),\textbf{b}(m)\}}(\eta) = \left\langle I \pm T_{\textbf{a}(m)+\textbf{b}(m)}\right\rangle_{\rho(m)}\geq 0$. This contradicts the induction assumption. Hence, case ii/b cannot occur.

Thus, irrespective of whether a given step in the circuit is a unitary transformation or a projective measurement, if the state after completing the step witnesses contextuality with the maximum negative value, so does the state before the step. By induction, the state before the first gate, i.e. the injected state, witnesses contextuality. $\Box$ 

\subsection{Coping with Mermin's square}

Mermin's square \cite{Merm} provides a beautifully simple proof of the Kochen-Specker theorem \cite{KS} in dimension four and higher, but for the programme of establishing contextuality of magic states as a quantum computational resource it poses a problem. Namely, the square can be converted into a contextuality witness of CSW type \cite{CSW} for which {\em{all}} two-qubit states come out contextual \cite{Howard}. But if contextuality is generic, then it is not a resource.

In more general terms,  Mermin's square exhibits the phenomenon of state-independent contextuality. It represents an obstacle to viewing contextuality as a resource possessed by some quantum states but not others.

When restricting to Pauli observables, state-in\-de\-pen\-dent contextuality only occurs in Hilbert spaces of even dimension \cite{Howard1}, and therefore was not an issue in \cite{Howard}.  However, in the present situation, the Hilbert space dimension is even, and furthermore, by a simple local rotation, Mermin's square can be embedded into real quantum mechanics.
\begin{equation}\label{RotM}
\parbox{3.1cm}{\includegraphics[width=3.1cm]{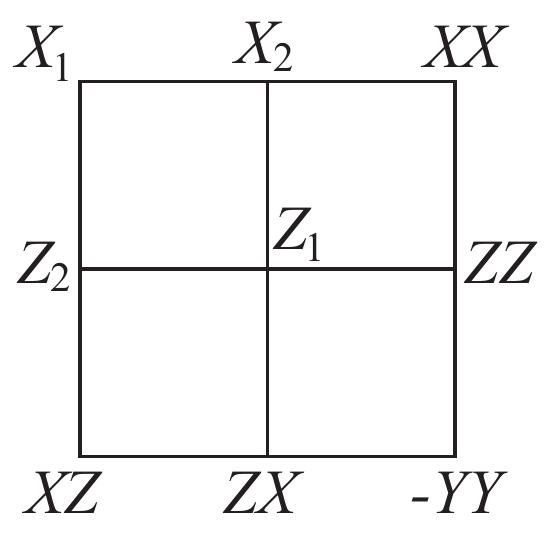}}
\end{equation}
Since, as in qudit QCSI, also in rebit QCSI contextuality is attributed to quantum states, state-independent contextuality seems likely to cause difficulty. Yet, in Theorem~\ref{T1} we established contextuality of magic states as a necessary resource. We thus have to explain why Mermin's square, and more generally the phenomenon of state-independent contextuality, did in fact not void  the contextuality-as-resource viewpoint.

To do so, we revisit the results established in Section~\ref{Cont}. First, by Theorem~\ref{NCC}, states with non-negative Wigner function are non-contextual. Hence contextuality is not generic, as required for a resource.

Next, we consider the rotated Mermin square, Eq.~(\ref{RotM}). For all columns and all rows except the bottom one, the belonging observables pairwise commute and generate a stabilizer group of CSS type. They can therefore be simultaneously measured in rebit QCSI. The measurement outcomes $\pm1$ must multiply to $+1$ in each of these contexts, which is implied by the identities among the observables, $X_1\cdot X_2 \cdot X_1X_2 = +I$ etc.

For the bottom row, the belonging observables $XZ$, $ZX$ and $-YY$ still commute and thus generate a stabilizer group, but this group is not of CSS type. As discussed at the beginning of Section~\ref{Cont}, these observables cannot be simultaneously measured in rebit QCSI. Therefore, the pre-determined measurement outcomes $\lambda(XZ)$, $\lambda(ZX)$, $\lambda(-YY)$ {\em{need not}} satisfy the constraint $\lambda(XZ) \lambda(ZX) \lambda(-YY) = -1$ implied by the operator relation $XZ \cdot ZX \cdot(-YY) = -I$. Therefore, $\lambda(\cdot) = +1$ for all observables in the rotated Mermin square is a consistent value assignment w.r.t. rebit QCSI. The algebraic contradiction vanishes because we have effectively removed the bottom row from the diagram (\ref{RotM}). 

This situation is handled by our definitions as follows: By Criterion~\ref{cr1}, $\{XZ,ZX,-YY\} \not \in {\cal{M}}$, c.f. Lemma~\ref{Mchar}. Therefore, $\lambda_\textbf{u}(XZ)\lambda_\textbf{u}(ZX)\lambda_\textbf{u}(-YY)=-1$ is not required by Definition~\ref{HVM1} of a non-contextual HVM (c.f. condition (ii)).

Generalizing the above observation, the phenomenon of state-independent contextuality does not come into play for the present setting of rebit QCSI, even if it does exist for systems of rebits. The reason is the restriction of the physical measurements to observables in ${\cal{O}}$, the set of pure-$X$ and pure-$Z$ Pauli operators.
\begin{Lemma}\label{lambdaHVM}
Consider a system of $n$ rebits where the measurable observables are restricted to the set ${\cal{O}}$. Then, the set ${\cal{S}}$ of consistent value assignments of a non-contextual HVM, $\lambda_\textbf{u}: {\cal{A}} \rightarrow \{\pm1\},\,\forall \textbf{u} \in {\cal{S}}$, is non-empty.
\end{Lemma}
Thus, there is no state-independent contextuality in rebit QCSI. Contextuality may persist at the level of probability.

{\em{Proof of Lemma~\ref{lambdaHVM}.}} The value assignments $\lambda_\textbf{u} : {\cal{A}} \longrightarrow \{\pm 1\}$, $\textbf{u} \in \mathbb{Z}_2^{2n}$, of Eq.~(\ref{ValAssignGen}) all satisfy the consistency condition Eq.~(\ref{coa}), $\lambda_\textbf{u}\left(T_{\textbf{a}+\textbf{b}}\right) = \lambda_\textbf{u}\left(T_{\textbf{a}}\right)\lambda_\textbf{u}\left(T_{\textbf{b}}\right)$, for all $T_\textbf{a},T_\textbf{b} \in {\cal{A}}$ such that  $[T_\textbf{a},T_\textbf{b}]=0$. By Lemma~\ref{Mchar}, for all $M \in {\cal{M}}$ and all $T_\textbf{a},T_\textbf{b} \in M$ ($[T_\textbf{a},T_\textbf{b}]=0$), it holds that $T_{\textbf{a}+\textbf{b}}=T_\textbf{a}T_\textbf{b}$, for all $\textbf{u} \in \mathbb{Z}_2^{2n}$. The value assignments $\lambda_\textbf{u}(\cdot)$ are thus consistent with the operator constraints. Hence, $\mathbb{Z}_2^{2n} \subseteq {\cal{S}}$, and ${\cal{S}} \neq \emptyset$. $\Box$\medskip

Finally, in Section~\ref{SecSCC} we argued that the contextuality witnesses ${\cal{W}}$ are closely related to state-independent contextuality, c.f. Remark 4. This is not in contradiction to the above statement that there is no state-independent contextuality in rebit QCSI. Namely, Remark 4 refers to rebit quantum mechanics (without the CSS restriction), not to rebit QCSI (with CSS restriction).

Let us revisit three facts from the preceding discussion. Assume two vectors $\textbf{a}, \textbf{b} \in V_{\cal{A}}$, $[\textbf{a},\textbf{b}]=0$, such that $T_{\textbf{a}+\textbf{b}} =-T_\textbf{a}T_\textbf{b}$.  Then, (i) (and only then) the witness ${\cal{W}}^{\{\textbf{a},\textbf{b}\}}$ can detect state-dependent contextuality. (ii) A state-independent contextuality proof can be constructed from  $T_{\textbf{a}+\textbf{b}}$, $T_\textbf{a}$, $T_\textbf{b}$ and local Pauli operators $X_i$, $Z_i$. (iii) The observables $T_\textbf{a}$, $T_\textbf{b}$ cannot be simultaneously measured in rebit QCSI (c.f. Lemma~\ref{Mchar}).

Now we note in addition that (iv) the states $\rho$ stabilized by such $\pm T_\textbf{a}$ and $\pm T_\textbf{b}$ make the witness ${\cal{W}}^{\{\textbf{a},\textbf{b}\}}_\rho(\textbf{x})$ maximally negative, i.e., ${\cal{W}}^{\{\textbf{a},\textbf{b}\}}_\rho(\textbf{x})=-2$ for suitable values of $\textbf{x}$. See Eq.~(\ref{wit}) for an example. The sets of commuting real Pauli operators which cannot be jointly measured  in rebit QCSI thus become stabilizer generators for states which maximally violate non-contextuality. The power of contextuality is transferred from measurement to (magic) states, exactly as it should be in a scheme of QCSI.

\subsection{CSS vs. real Clifford transformations}

It is instructive to examine what happens if the restricted gate set of rebit QCSI is extended from the CSS-ness preserving Clifford operations to the lager set of real Clifford operations. This comprises, in particular, increasing the set of physically measurable observables from ${\cal{O}}$ to the larger set ${\cal{A}}$ of all real Pauli observables. 

This change opens the door to state-independent contextuality; See Mermin's square in Eq.~(\ref{RotM}). As discussed in the previous section, this is not compatible with viewing contextuality as a resource possessed only by special quantum states. 

As for the Wigner function in relation to contextuality, for $n\geq2$ rebits, a positive
Wigner function no longer yields a non-contextual hidden variable model.
Namely, with the new measurement contexts available, the ``pre-determined''
measurement outcomes $\lambda_\textbf{u}$ assigned by the Wigner function via Eq.~(\ref{ValAssign}) to HVM states $\textbf{u} \in V$ fail to satisfy Eq.~(\ref{Valco}) in  
Definition~\ref{HVM1} of a non-contextual HVM.

As for the Wigner function in relation to efficient sampling, (i) $W$ produces the correct
quantum mechanical expectation values for all observables in ${\cal{A}}$ via Eq.~(\ref{Trip}). Hence, it also produces the correct expectation values for all observables defined for real states. (ii) If $W_\rho>0$, then for all $A \in {\cal{A}}$ the expectation values $\langle A \rangle_\rho$ 
can be efficiently estimated by sampling.  However, (iii) Allowing measurements in the middle of the computation, those in ${\cal{A}}\backslash {\cal{O}}$ can introduce negativity into the Wigner function, preventing efficient sampling from it.

To summarize, as an example for tinkering with rebit QCSI, if the restricted gate set of CSS-ness preserving Clifford operations is replaced by the broader class of real Clifford operations, then contextuality is undone as a resource, the link between positive Wigner functions and non-contextual hidden variable models breaks, and efficient classical simulation, by sampling from the Wigner function, of rebit QCSI without magic states is obstructed (other simulation methods \cite{DG}, \cite{BVN} remain, though).  These observations  illustrate the intricate relation among the various constituents of rebit QCSI.

\section{Conclusion}
\label{Concl}

We have established that contextuality and Wigner function negativity are necessary resources for  computational universality of the discussed scheme of quantum computation by state injection on rebits. To this end, we have constructed the computational scheme itself, and supplemented it with a matching Wigner function (complete with a Hudson's theorem and efficient sampling algorithm for positive Wigner functions) and contextuality witnesses. These parts mutually reinforce each other: Efficient sampling provides operational justification for calling states with positive Wigner function ``classical'', and Hudson's theorem ensures that the notions of classicality established by the Wigner function and by the gate restrictions in rebit QCSI match. The absence of Wigner function negativity and contextuality reveal the limitations of the restricted gate set. Furthermore, our computational scheme is constructed  in such a way that state-independent contextuality does not come into play, even if it  is present in rebits.

We have thus extended all the essential properties that held for QCSI in the case of qudits of odd prime dimension \cite{Howard}, \cite{Veitch} to rebits, with the sole exception that for the present rebit scheme, Wigner function negativity does not imply contextuality.\medskip  

To widen the scope of the discussion, we note that contextuality has also been established as a resource for measurement-based quantum computation (MBQC) \cite{AB}, \cite{Hoban}, \cite{RR1}. Namely, in MBQC contextuality is necessary for the ability to compute non-linear Boolean functions. One may thus want to compare the roles played by contextuality in QCSI and MBQC. But there was an obstacle: The MBQC result has to date only been established for the case of 2-level systems, where most of the existing results \cite{Howard} do not apply. The present paper removes this mismatch, and thus prepares the ground for a comparison between the two computational schemes.

We conclude with three open questions.
\begin{itemize}
\item{In QCSI contextuality is about speedup, as one might expect for a scheme of quantum computation. But in MBQC it is about computability. What is the reason for this dichotomy?}
\item{Due to the formulation in terms of a Wigner function, covariance plays an important role for QCSI. Is covariance also a useful concept in the discussion of MBQC?}
\item{We noted that the restricted gate set in the present scheme of rebit QCSI is precisely the gate set that can be implemented by defect braiding and fusion with surface codes \cite{RH07}. There is more complicated lattice surgery by which, in addition, the Hadamard gate can be realized  \cite{Bombin}, \cite{Fowler}. In this way, the full real subgroup of the qubit stabilizer group becomes available as the restricted gate set, and the real stabilizer states are the ``cheap'' / non-magic states. Is there a Wigner function with matching Hudson's theorem and covariance property?}
\end{itemize}

{\em{Acknowledgments:}} RR thanks Rob Spekkens for discussion. ND is supported by the Lockheed Martin Corporation. PAG is supported by FRQNT (Quebec), and RR is supported by NSERC, Cifar and IARPA.

\appendix

\section{Fourier transform on the group $\Z_2^n$} \label{section:fourier}

The Fourier transform of a function $f :M \rightarrow \R$ defined on a linear subspace $M$ of $\Z_2^n$ is the function ${\cal F} f$ or $\hat f$ defined by
$$
{\cal F} f({\bf u}) = \frac{1}{\sqrt{|M|}} \sum_{{\bf x} \in M} (-1)^{({\bf u}, {\bf x})} f({\bf x}).
$$

\begin{Lemma} \label{lemma:fourier_involutive}
The Fourier transform $\cal F$ is involutive, \emph{i.e.} ${\cal F} \circ {\cal F} = Id$, or equivalently $\cal F$ is its own inverse. 
\end{Lemma}

{\em{Proof of Lemma~\ref{lemma:fourier_involutive}.}} Let us determine the image of a function $f: M \rightarrow \R$ by $\cal F \circ \cal F$.
\begin{align*}
\left( ({\cal F} \circ {\cal F}) (f) \right) ({\bf u})
& = {\cal F} \left( {\cal F}(f) \right) ({\bf u})\\
& = \frac{1}{\sqrt{|M|}} \sum_{{\bf x} \in M} (-1)^{({\bf u, x})} {\cal F}f({\bf x})\\
& = \frac{1}{|M|} \sum_{{\bf x} \in M} \sum_{{\bf y} \in M} (-1)^{({\bf u, x})} (-1)^{({\bf x, y})} f({\bf y})\\
& = \frac{1}{|M|} \sum_{{\bf y} \in M} \left( \sum_{{\bf x} \in M} (-1)^{({\bf x, u+y})} \right) f({\bf y})\\
& = \sum_{{\bf y} \in M} \delta_{\bf u, y} f({\bf y})\\
& = f({\bf u}),
\end{align*}
which demonstrates the claim. $\Box$

\section{Properties of the rebit Wigner function} \label{WiProp}

\begin{Lemma} \label{lemma:A_orthonormal}
The set of Pauli operators $\mathcal{A}$ is an orthonormal basis of the space $S_{2^n}(\R)$ of symmetric matrices of size $2^n$ endowed with the inner product $(A, B) = \frac{1}{2^n}\Tr(A^T B)$. 
\end{Lemma}

{\em{Proof of Lemma~\ref{lemma:A_orthonormal}.}} Denote by $E_{i, j}$ the matrix with entry 0 everywhere expect at the intersection of the $i$-th row and $j$-th column where it is 1. The space $S_N(\R)$ is generated by the matrices $(E_{i, j}+E_{j, i})$ with $1 \leq i<j \leq N$ and $E_{i, i}$ with $1 \leq i \leq N$. Moreover, we can easily check that these  $N(N+1)/2$ matrices are independent. Thus the dimension of $S_N(\R)$ is $N(N+1)/2$. In our case $N=2^n$ and $\dim S_{2^n}(\R) = 2^{2n-1} + 2^{n-1}$.

The set $\cal A$ contains $2^{2n-1} + 2^{n-1}$ symmetric matrices. These matrices are pairwise orthogonal, \emph{i.e.} they satisfy $\frac{1}{2^n}\Tr(T_u T_v) = 0$ when $u \neq v$, thus they are linearly independent. This proves that they form a basis of the space of symmetric matrices. The orthonormality is a consequence of the orthonormality of the Paulis. $\Box$ \medskip

We show that from a given Wigner function we can obtain the corresponding real density operator, proving that the Wigner function is informationally complete.
\begin{Lemma} \label{lemma:wigner_invertible}
Let $\rho$ be a real density operator and let $W_\rho$ be its Wigner function. Then $\rho$ satisfies
$$
\rho = \sum_{{\bf u} \in \Z_2^{2n}} W_\rho({\bf u}) A_{\bf u}.
$$
\end{Lemma}

{\em{Proof of Lemma~\ref{lemma:wigner_invertible}.}} We expand the r.h.s. of the above equation by inserting the definition Eq.~(\ref{WFa}) of $W_\rho$ and Eq.~(\ref{eqn:A_u_second_form}) for $A_\textbf{u}$, and obtain
\begin{align*}
\sum_{{\bf u} \in \Z_2^{2n}} W_\rho({\bf u}) A_{\bf u}
& =  \frac{1}{2^{3n}} \sum_{\substack{{\bf u} \in \Z_2^{2n} \\ {\bf v, w} \in V_{\cal A}}} (-1)^{[{\bf u}, {\bf v + w}]} \Tr(T_{\bf v} \rho) T_{\bf w}
\end{align*}
Now note that the sum $\sum_{{\bf u} \in \Z_2^{2n}} (-1)^{[{\bf u}, {\bf v}+{\bf w}]}$ is $2^{2n} \delta_{v, w}$, which is a standard property of characters. Hence,
$$
\sum_{{\bf u} \in \Z_2^{2n}} W_\rho({\bf u}) A_{\bf u} = \frac{1}{2^n} \sum_{{\bf w} \in V_{\cal A}} \Tr(T_{\bf w} \rho) T_{\bf w}.
$$
From Lemma~\ref{lemma:A_orthonormal}, this sum is the decomposition of $\rho$ in the orthonormal basis ${\cal A}$. This proves that we recover the state $\rho$. $\Box$ \medskip

{\em{Proof of Property~4 (Section~\ref{Widef}).}} With the definition Eq.~(\ref{mwf}),
$$
\begin{array}{rcl}
A_0 &=& \frac{1}{2^n} \sum_{\textbf{u} \in V |\, (\textbf{u}_X,\textbf{u}_Z) \!\!\! \mod 2=0} T_\textbf{u}\vspace{2mm}\\
&=& \frac{1}{2^{n+1}} \big(\prod_{i=1}^n (I+Z_i) \prod_{j=1}^n (1+X_j) + \vspace{1mm}\\
&& + \prod_{i=1}^n (I+X_i) \prod_{j=1}^n (1+Z_j) \big) \vspace{2mm}\\
& =& 2^{n-1} \left( |0_n\rangle \langle 0_n| |+_n\rangle \langle +_n| +  |+_n\rangle \langle +_n| |0_n\rangle \langle 0_n|\right),
\end{array}
$$
and thus
\begin{equation}\label{A0}
A_0 = 2^{\frac{n}{2} -1} \left( |0_n\rangle \langle +_n| +  |+_n\rangle \langle 0_n|\right).
\end{equation}
Further using the properties that $\rho$ is Hermitian and real,
$$
W_\rho(\textbf{v}) = \frac{1}{\sqrt{2}^n} \langle 0_n| T_\textbf{v}^\dagger \rho T_\textbf{v} |+_n\rangle.
$$
Now, we consider the case where $\rho_{AB}$ factorizes, $\rho_{AB}=\sigma_A \otimes \tau_B$. We may write any phase space point $\textbf{v}$ as $\textbf{v} =\textbf{v}_A + \textbf{v}_B$, where $\textbf{v}_A$ ($\textbf{v}_B$) acts non-trivially only on system $A$ ($B$). Then, $T_\textbf{v} =\pm T_{\textbf{v}_A} T_{\textbf{v}_B}$ and
$$
\begin{array}{rcl}
W_{\sigma \otimes \tau}(\textbf{v}) &=&  \displaystyle{\frac{\langle 0_{n_A},\!0_{n_B}| T_{\textbf{v}_B}^\dagger T_{\textbf{v}_A}^\dagger \sigma\otimes \tau \,T_{\textbf{v}_A} T_{\textbf{v}_B}|+_{n_A},\!+_{n_B}\rangle}{\sqrt{2}^{n_A+n_B}} }\vspace{1mm}\\
&=& W_\sigma(\textbf{v}_A) W_\tau(\textbf{v}_B).
\end{array}
$$\medskip

{\em{Proof of Property~5 (Section~\ref{Widef}).}} We define modified phase point operators
$$
\tilde{A}_\textbf{u} = \sum_{\textbf{v} \in {\cal{A}}} (-1)^{[\textbf{u},\textbf{v}]} T_\textbf{v} + i \sum_{\textbf{v} \in {\cal{T}}\backslash {\cal{A}}} (-1)^{[\textbf{u},\textbf{v}]} T_\textbf{v},
$$ 
such that 
\begin{equation}\label{AtilOrtho}
\text{Tr} (\tilde{A}_\textbf{u} \tilde{A}_\textbf{v}) = 2^n \delta_{\textbf{u},\textbf{v}}.
\end{equation}
For a real Hermitian operator $\rho$ we therefore have
$$
\begin{array}{rcl}
\rho &=& \displaystyle{\frac{1}{2^n} \sum_{\textbf{v} \in \mathbb{Z}_2^{2n}} \text{Tr} (\tilde{A}_\textbf{v}\rho) \tilde{A}_\textbf{v} =  \frac{1}{2^n} \sum_{\textbf{v} \in \mathbb{Z}_2^{2n}} \text{Tr} (A_\textbf{v}\rho) \tilde{A}_\textbf{v}}\\
&=& \displaystyle{\sum_{\textbf{v}\in \mathbb{Z}_2^{2n}} W_\rho(\textbf{v}) \tilde{A}_\textbf{v}.}
\end{array}
$$
The second equality holds because $\rho$ is real. Using the last expression together with Eq.~(\ref{AtilOrtho}), we find for two real Hermitian operators $\rho$, $\sigma$ that
$$
\text{Tr}(\rho\sigma) = 2^n \sum_{\textbf{v} \in \mathbb{Z}_2^{2n}} W_\rho(\textbf{v})W_\sigma(\textbf{v}),
$$
as claimed. $\Box$\medskip

{\em Proof of Lemma~\ref{lemma:hudson_expression_W}.} By inserting into the Wigner function the expansion 
$\proj \psi = \sum_{\bf x, \bf y \in \Z_2^n} \psi({\bf x}) \psi({\bf y})\ketbra{\bf x}{\bf y}$, we obtain
$$
W_\psi({\bf u}) = \frac{1}{2^n} \sum_{\bf x,  y \in \Z_2^n} \psi({\bf x})\psi({\bf y}) \Tr(T_{\bf u} A_0 T_{\bf u}^\dagger \ketbra{\bf x}{\bf y}).
$$
We use the expression Eq.~(\ref{A0}) for $A_0$, and verify by direct calculation that 
\begin{equation} \label{eqn:appendix_W_computation1}
\Tr(T_{\bf u} \ket{0_n} \bra{+_n} T_{\bf u}^\dagger \ketbra{{\bf x}}{{\bf y}})
= \frac{(-1)^{({\bf x} + {\bf u}_X, {\bf u}_Z)}}{2^{n/2}} \delta_{{\bf u}_X, {\bf y}}.
\end{equation}
Inserting Eq.(\ref{eqn:appendix_W_computation1}) in the above $W_\psi(\textbf{u})$ gives
\begin{align*}
W_\psi({\bf u})
	& = \frac{1}{2^{n+1}} 
		\left( \sum_{\bf x \in \Z_2^n}
			\psi({\bf x})\psi({\bf u}_X) (-1)^{({\bf x} + {\bf u}_X, {\bf u}_Z)} 
		\right)\\
	&	+ \frac{1}{2^{n+1}} 
		\left( \sum_{\bf y \in \Z_2^n} 
			\psi({\bf u}_X)\psi({\bf y}) (-1)^{({\bf y+u}_X, {\bf u}_Z)} 
		\right)\\
	& = \frac{1}{2^n} \sum_{\bf x \in \Z_2^n} (-1)^{({\bf x}, {\bf u}_Z)} \psi({\bf u}_X) \psi({\bf u}_X + {\bf x}).
\end{align*}
as stated in Lemma~\ref{lemma:hudson_expression_W}.
$\Box$

\section{CSS-ness preserving Clifford gates} \label{appsection:CSS_Clifford}

Here, we prove Lemma~\ref{appcor:G_CSS_generators} from Section \ref{CSS}, and Lemmas~\ref{prop:conjugation_discretization}, \ref{prop:CSS_morphism} from Section~\ref{section:covariance}, about the structure of the CSS-ness preserving subgroup of the Clifford group.\medskip

{\em{Proof of Lemma~\ref{prop:conjugation_discretization}.} } First, we omit the phase $\pm 1$ of $T_{\bf a}$ and focus on the effect of the conjugation on the vector~${\bf a}$.
Let $g \in G_{CSS}$ and let $\vphi_g$ be the automorphism of the real Pauli group $P_n(\mathbb{R})$ defined by conjugation by $g$:
\begin{equation}\label{phig}
\begin{array}{rl}
\vphi_g: P_n(\mathbb{R}) & \longrightarrow P_n(\mathbb{R})\vspace{1mm}\\
Q & \longmapsto g Q g^{\dagger}\\
\end{array}
\end{equation}
This morphism of  $P_n(\mathbb{R})$ induces a morphism of its quotient $P_n(\mathbb{R})/\{\pm I\}$, which is isomorphic to $\Z_2^{2n}$, that is $\vphi_g$ induces a matrix $F \in M_{2n}(\Z_2)$ such that
$$
g T_{\bf a} g^\dagger = \lambda({\bf a}) T_{F{\bf a}},
$$
where $\lambda({\bf a}) \in \{\pm 1\}$.
Since $\vphi_g$ is an automorphism, $F \in \GL_{2n}(\Z_2)$.
Moreover, the conjugation preserves the commutation relation and we know that $T_{\bf a}$ and $T_{\bf b}$ commute if and only if $[{\bf a}, {\bf b}] = 0$. This proves that $F \in \Sp_{2n}(\Z_2)$. $\Box$\medskip

{\em{Proof of Lemma~\ref{prop:CSS_morphism}.}} Consider a pair $g, g' \in G_{CSS}$ and denote by ${\cal F}(g) = A(F, {\bf t})$ and ${\cal F}(g') = A(F', {\bf t'})$ their images. The value of ${\cal F}(gg')$ is defined by the conjugation by $gg'$. We obtain
\begin{align*}
gg' T_{\bf a} (gg')^\dagger & = g \left(g' T_{\bf a} g'^\dagger \right) g^\dagger\\
& = g \left( (-1)^{[{\bf t'}, F'{\bf a}]} T_{F'{\bf a}} \right) g^\dagger\\
& = (-1)^{[{\bf t'}, F'{\bf a}]+[{\bf t}, FF'{\bf a}]} T_{FF'{\bf a}}\\
& = (-1)^{[F{\bf t'}+{\bf t}, FF'{\bf a}]} T_{FF'{\bf a}}
\end{align*}
Therein, we have used $[{\bf t'}, F'{\bf a}] = [F{\bf t'}, FF'{\bf a}]$.
This gives ${\cal F}(gg') = A(FF', F{\bf t'}+{\bf t})$, which is indeed the composition of ${\cal F}(g) = A(F, {\bf t})$ and ${\cal F}(g') = A(F', {\bf t'})$. Hence, ${\cal F}(gg') = {\cal{F}}(g){\cal{F}}(g')$, for all $g,g' \in G_{CSS}$. $\Box$\medskip

{\em{Proof of Lemma~\ref{appcor:G_CSS_generators}.}} Our fist goal is to describe $G_{CSS}$ as the normalizer of the special Pauli operators ${\cal{O}}$.

\begin{Lemma} \label{applemma:clifford_X_Z}
The group $G_{CSS}$ is the normalizer in $O_{2^n}(\R)$ of the set ${\cal{O}} = \{Z({\bf u}) \ | \ {\bf u} \in \Z_2^n\} \cup \{X({\bf v}) \ | \ {\bf v} \in \Z_2^n\}$ of Pauli-observables which have only an $X$-part or only a $Z$-part.
\end{Lemma}

{\em{Proof of Lemma~\ref{applemma:clifford_X_Z}.}} If $g$ belongs to the normalizer of $\cal O$, then it conserves CSS codes and CSS states.

In order to obtain the inverse implication, we will show that an operator $g$ which preserves CSS states, stabilizes the set of all CSS groups by conjugation. Applying this argument to rank one groups $\langle X_i\rangle$ and $\langle Z_i\rangle$, we obtain the lemma.
Thus, we want to prove that the image under conjugation by $g$ of a CSS group is also a CSS group. 
This is true when $S$ has rank $n$. We work by induction.
Assume the result for every CSS group of rank $r$ and let us prove that it is also true for a CSS group $S$ of rank $r-1$.
Let $S'$ be the group $gSg^\dagger$ obtained after conjugation and let $M$ be the subspace $$M = \{ {\bf a} \in \Z_2^{2n} \ | \ \pm T_{\bf a} \in S'\}.$$
We associate with $M$ two subspaces 
$$
M_Z = \{ {\bf u}_Z \in \Z_2^n \ | \ \exists \ ({\bf u}_Z, {\bf u}_X) \in M \}$$
and 
$$
M_X = \{ {\bf u}_X \in \Z_2^n \ | \ \exists \ ({\bf u}_Z, {\bf u}_X) \in M \}.
$$
Note that $M \subset M_Z \oplus M_Z$ and $S'$ is a CSS code if and only if we have equality $M = M_Z \oplus M_Z$ and in that case $M_Z$ and $M_X$ are two orthogonal subspaces.
Assume that $S'$ is not a CSS code then $M_Z \oplus M_X$ contains strictly $M$ and has dimension
$$
\dim M_Z \oplus M_X > \dim M = r-1.
$$
Now, choose two logical operators $\bar X$ and $\bar Z$ for the code $S$ which anti-commute. The two CSS groups $\langle S, \bar X\rangle$ and $\langle S, \bar Z\rangle$ are sent onto CSS groups by conjugation. Denote by $N$ and $R$ respectively the corresponding subspaces of $\Z_2^{2n}$, defined as $M$. These two spaces can be decomposed as 
$$
N = N_Z \overset{\perp}{\oplus} N_X \text{ and } R = R_Z \overset{\perp}{\oplus} R_X.
$$
These spaces both contain $M_Z \oplus M_X$ and have dimension $r$, hence we have $M_Z \oplus M_X = N = R$.
To find a contradiction, consider the operators $g \bar X g^\dagger = \lambda({\bf a}) T_{\bf a}$ and $g \bar Z g^\dagger = \lambda({\bf b}) T_{\bf b}$. By construction, we have ${\bf a} \in N$ and ${\bf b} \in R$. Using the equality $N=R$, we can see that the two inner products $({\bf a}_Z, {\bf b}_X)$ and $({\bf b}_Z, {\bf a}_X)$ are 0, which implies that $T_{\bf a}$ and $T_{\bf b}$ commute. This is a contradiction since $g$ preserves the commutation relation.
Finally, we proved that $S'$ is a CSS group. The set of all CSS group is preserved by conjugation by $g$. $\Box$ 
\medskip

We now return to the subject of Lemma~\ref{prop:conjugation_discretization}, and further characterize the matrices $F$ appearing on the r.h.s. of Eq.~(\ref{gTg}). These matrices have one of the two following block structures.
\begin{align} \label{eq:F_CSS}
F =
\begin{pmatrix}
F_Z & 0\\
0 & F_X\\
\end{pmatrix}
\quad \text{ or } \quad
F =
\begin{pmatrix}
0 & F_X\\
F_Z & 0\\
\end{pmatrix}
\end{align}
where $F_Z, F_X \in \GL_{n}(\Z_2)$ and $F_X = (F_Z^{-1})^t$.
In what follows, we denote by $F_{CSS}$ the set of symplectic matrices introduced in Eq.~(\ref{eq:F_CSS}).
The result is that every CSS Clifford operator induces a pair $(F, {\bf x}) \in F_{CSS} \times \Z_2^{2n}$. 
\medskip

To demonstrate Eq.~(\ref{eq:F_CSS}), note that the conjugation $\vphi_g$ of Eq.~(\ref{phig}) preserves the set of CSS operators $X({\bf u})$ and $Z({\bf v})$. Suppose an operator $X({\bf u})$ is sent onto $X({{\bf u}'})$ and that $X({\bf v})$ is sent onto $Z({{\bf v}'})$. Then, the image of $X({{\bf u}+{\bf v}})$ is $X({{\bf u}'})Z({{\bf v}'})$ which is impossible. Therefore, $\vphi_g$ has two possible structures, either it conserves both sets $\{X({\bf u}) \ | \ {\bf u} \in \Z_2^n \}$ and $\{Z({\bf u}) \ | \ {\bf u} \in \Z_2^n \}$, or it exchanges these two sets. 
This proves that the matrix $F$ has one of the two following block structures.
$$
F =
\begin{pmatrix}
F_Z & 0\\
0 & F_X\\
\end{pmatrix}
\quad \text{ or } \quad
F =
\begin{pmatrix}
0 & F_X\\
F_Z & 0\\
\end{pmatrix}
$$
where $F_Z, F_X \in \GL_{n}(\Z_2)$. Finally, $F_Z = (F_X^t)^{-1}$ is a consequence of the requirement that the $F$'s must preserve the symplectic form.

\medskip
The knowledge of the structure of the matrix $F$ will now be useful to determine the phase $\lambda({\bf a})$ of the operator $g T_{\bf a} g^\dagger = \lambda({\bf a}) T_{F{\bf a}}$.
Since every character of $\Z_2^{2n}$ is of the form ${\bf a} \mapsto (-1)^{\bf [x, a]}$ for some vector ${\bf x}$ of $\Z_2^{2n}$, it suffices to show that $\lambda$ is the restriction of such a character to the set $V_{\cal A}$. Denote by $({\bf e}_i)_{i=1}^{2n}$ the canonical basis of the space $\Z_2^{2n}$ and denote by $\mu$ the character of $\Z_2^{2n}$ defined by $\mu({\bf e}_i) = \lambda({\bf e}_i)$. To prove that $\mu = \lambda$ on the set $V_{\cal A}$, it is enough to show that
\begin{itemize}
\item If ${\bf a}_X ={\bf b}_X=0$ or if ${\bf a}_Z = {\bf b}_Z = 0$, then we have $\lambda({\bf a}+{\bf b}) = \lambda({\bf a})\lambda({\bf b})$.
\item If ${\bf a} = ({\bf a}_Z, {\bf a}_X) \in V_{\cal A}$, then we have $\lambda({\bf a}) = \lambda(({\bf a}_Z, 0))\lambda((0, {\bf a}_X))$.
\end{itemize}
In what follows, we assume that $F$ is block diagonal. The proof is similar in the anti-diagonal case.
If ${\bf a}_X = {\bf b}_X = 0$, then we have $\varphi_g(T_{{\bf a}+{\bf b}}) = \lambda({\bf a}+{\bf b})T_{F({\bf a}+{\bf b})}$ which is also $\varphi_g(T_{\bf a}T_{\bf b}) = \lambda({\bf a})\lambda({\bf b})T_{(F_Z {\bf a}_Z, 0)}T_{(F_Z {\bf b}_Z, 0)} =  \lambda({\bf a})\lambda({\bf b}) T_{F({\bf a}+{\bf b})}$. The equality $\lambda({\bf a}+{\bf b}) = \lambda({\bf a})\lambda({\bf b})$ follows.
The proof of the second implication is similar.

This implies that $\lambda$ coincides with the character $\mu$ on the set $V_{\cal A}$, which means that $\lambda({\bf a}) = (-1)^{[{\bf x}, {\bf a}]}$ for some vector ${\bf x} \in \Z_2^{2n}$. \medskip

To illustrate the above with examples, we list the pairs $(F, {\bf x})$ for a few gates  $g \in G_{CSS}$  of special interest.
\begin{itemize}
\item \label{appitem:hadamard} If $g = \otimes_i H_i$ then
\begin{equation} \label{appeqn:F_hadamard}
F = 
\begin{pmatrix}
0 & I_n\\
I_n & 0
\end{pmatrix}
\text{ and }
{\bf x} = 0.
\end{equation}

\item If $g = CNOT(i, j)$ then
\begin{equation} \label{appeqn:F_CNOT}
F = 
\begin{pmatrix}
I_n + E_{i, j} & 0\\
0 & I_n + E_{j, i}
\end{pmatrix}
\text{ and }
{\bf x} = 0.
\end{equation}
The matrix $E_{i, j}$ denote the $n \times n$ binary matrix whose only non-zero coefficient is in position $(i, j)$
\item If $g = T_{\bf u}$ then
\begin{equation} \label{appeqn:F_Pauli}
F = I_n
\quad \text{ and } \quad
{\bf x} = {\bf u}.
\end{equation}
\end{itemize}
The fact that the pair $(F, {\bf x})$ associated with a Pauli operator $T_{\bf u}$ is $(0, {\bf u})$ is a direct consequence of the commutation relations between Pauli operators.\medskip

We now return to the subject of Lemma~\ref{prop:CSS_morphism}, and describe the image of the map ${\cal{F}}$. It holds that
$$
\im {\cal F} = \{ A(F, {\bf t}) \ | \ F \in F_{CSS}, {\bf t} \in \Z_2^{2n} \}.
$$
Recall that this application $\cal F$ is well defined by unicity in Lemma~\ref{prop:conjugation_discretization}.
The translation vector ${\bf t}$ and the vector ${\bf x}$ of Lemma~\ref{prop:conjugation_discretization} are related by the equation ${\bf t} = F {\bf x}$.

In order to determine the image of $\cal F$, note that we already know some elements of $\im \cal F$. Indeed, from Eq.~(\ref{appeqn:F_CNOT}), all the transformations $A(F, 0)$ with
$$
F = 
\begin{pmatrix}
I_n + E_{i, j} & 0\\
0 & I_n + E_{j, i}
\end{pmatrix}
$$
belong to this subgroup.
The matrices $I_n + E_{i, j}$ are called transvection matrices and they are known to generate the group $\SL_n(\Z_2)$, which coincides with $\GL_n(\Z_2)$. This implies that $\im \cal F$ contains all the $A(F, 0)$ associated with
\begin{equation} \label{appeqn:imF}
F = 
\begin{pmatrix}
M & 0\\
0 & (M^{-1})^t
\end{pmatrix},
\end{equation}
where $M \in \GL_n(\Z_2)$. 

This means that $\im \cal F$ contains all the block diagonal matrices of $F_{CSS}$. The anti-diagonal matrices of $F_{CSS}$ can be obtained by multiplication with the matrix $F$ of Eq.~(\ref{appeqn:F_hadamard}). This shows that $\im \cal F$ contains all the affine maps $A(F, 0)$, with $F \in F_{CSS}$. Finally, to reach $A(F, {\bf t}) = A(0, {\bf t}) A(F, 0)$, note that $A(0, {\bf t}) \in \im \cal F$ by Eq.~(\ref{appeqn:F_Pauli}). $\Box$ \medskip

Thanks to this group morphism, we obtain a complete description of the group $G_{CSS}$. First, we have the group isomorphism
\begin{equation} \label{appeqn:G_CSS_isomorphism}
G_{CSS} / \Ker {\cal F} \simeq \im {\cal F}.
\end{equation}
By construction of $\cal F$, its kernel is the set of orthogonal matrices commuting with every matrix. This is $\{\pm I_{2^n}\}$. We have seen above that $\im \cal F$ is generated by the images of $\otimes_i H_i, CNOT(i, j)$ and $T_{\bf u}$. Thus, from the previous isomorphism, the group $G_{CSS}$ is generated by these 3 types of operators and by $\Ker({\cal F}) = \{\pm I\}$. This proves the first part of Lemma~\ref{appcor:G_CSS_generators}.

The Affine group $\AGL(\Z_2)$ is known to be the semi-direct product of the group of translations by the general linear group. We obtain a similar structure for $\im \cal F$. It is the semi-direct product of the group of translations $A(0, {\bf t})$ by $F_{CSS}$, which implies
\begin{equation} \label{appeqn:imF_sdproduct}
\im {\cal F} \simeq \Z_2^{2n} \rtimes F_{CSS}.
\end{equation}
To prove this decomposition, it is suficient to check that these two sets are subgroups of $\im \cal F$ which jointly generate $\im \cal F$ and that the subgroup of translations is a normal subgroup.

By definition the group $F_{CSS}$ can also be decomposed as a semi-direct product
\begin{equation} \label{appeqn:F_CSS_sdproduct}
F_{CSS} \simeq \GL_n(\Z_2) \rtimes \Z_2.
\end{equation}
The set of block-diagonal matrices of $F_{CSS}$ is a subgroup isomorphic to $\GL_n(\Z_2)$ and it is normal since it is a subgroup of index 2 of $F_{CSS}$. The second component is the subgroup of $F_{CSS}$ generated by the matrix $F$ of Eq.~(\ref{appeqn:F_hadamard}) of order 2, and is isomorphic to $\Z_2$.
The second item of Lemma~\ref{appcor:G_CSS_generators} follows from Eqs.~(\ref{appeqn:G_CSS_isomorphism}), (\ref{appeqn:imF_sdproduct}) and (\ref{appeqn:F_CSS_sdproduct}). $\Box$


\begin{thebibliography}{99}

\bibitem{Magic}
S. Bravyi and A. Kitaev, {\em{Universal quantum computation with ideal Clifford gates and noisy ancillas}}, Phys. Rev. A \textbf{71}, 022316 (2005).

\bibitem{Bell}
J.S. Bell, {\em{On the Problem of Hidden Variables in Quantum Mechanics}}, Rev. Mod. Phys. \textbf{38}, 447 (1966).

\bibitem{KS}
S. Kochen and E.P. Specker, {\em{The Problem of Hidden Variables in Quantum Mechanics}}, J. Math. Mech. \textbf{17}, 59 (1967).

\bibitem{Abram}
Samson Abramsky, Adam Brandenburger, {\em{The sheaf-theoretic structure of non-locality and contextuality}}, New J. Phys. \textbf{13}, 113036 (2011).

\bibitem{Acin}
A. Ac{\'i}n, T. Fritz, A. Leverrier, A. Bel{\'e}n Sainz, {\em{A Combinatorial Approach to Nonlocality and Contextuality}}, arXiv:1212.4084.

\bibitem{Howard}
M. Howard {\em{et al.}}, {\em{Contextuality supplies the ÔmagicÕ for quantum computation}}, Nature \textbf{510}, 351355 (2014).

\bibitem{Galv1}
Ernesto F, Galv{\~a}o, {\em{Discrete Wigner functions and quantum computational speedup}}, Phys. Rev. A \textbf{71}, 042302 (2005).

\bibitem{Veitch}
V. Veitch, C. Ferrie, D. Gross and J. Emerson, {\em{Negative quasi-probability as a resource for quantum computation}}, New J. Phys. \textbf{14}, 113011 (2012).

\bibitem{Wigner}
E. Wigner, {\em{On the Quantum Correction For Thermodynamic Equilibrium}}, Phys. Rev. \textbf{40}, 749 (1932).

\bibitem{Gibbons}
K.S. Gibbons, M.J. Hoffman, and W.K. Wootters, {\em{Discrete Phase Space Based on Finite Fields}}, Phys. Rev. A \textbf{70}, 062101 (2004).

\bibitem{Gross}
D. Gross, {\em{Computational power of quantum many-body states and some results on discrete phase spaces}}, PhD Thesis, Imperial College London, 2005.

\bibitem{Spekkens}
R.W. Spekkens, {\em{Negativity and Contextuality are Equivalent Notions of Nonclassicality}}, Phys. Rev. Lett. \textbf{101}, 020401 (2008).

\bibitem{DG}
D. Gottesman, {\em{Stabilizer Codes and Quantum Error Correction}}, Ph.D. thesis, California Institute of Technology, 1997.

\bibitem{Mari}
A. Mari, J. Eisert, {\em{Positive Wigner functions render classical simulation of quantum computation efficient}}, Phys. Rev. Lett. \textbf{109}, 230503 (2012).

\bibitem{Galv2}
Cecilia Cormick, Ernesto F. Galv{\~a}o, Daniel Gottesman, Juan Pablo Paz, and Arthur O. Pittenger, {\em{Classicality in discrete Wigner functions}}, Phys. Rev A \textbf{73}, 012301 (2006).

\bibitem{Merm}
 N.D. Mermin, {\em{Hidden variables and the two theorems of John Bell}}, Rev. Mod. Phys. {\bf{65}}, 803 (1993).
 
\bibitem{rebit_rudolph}
T. Rudolph, L. Grover,  {\em{A 2 rebit gate universal for quantum computing}}, arXiv:quant-ph/0210187.

\bibitem{CSS}
A.R. Calderbank, E.M. Rains, P.W. Shor, N.J.A. Sloane, {\em{Quantum Error Correction and Orthogonal Geometry}}, Phys. Rev. Lett. \textbf{78}, 405 (1997).

\bibitem{CSW}
Adan Cabello, Simone Severini, Andreas Winter, {\em{Graph-Theoretic Approach to Quantum Correlations}}, Phys. Rev. Lett. \textbf{112}, 040401 (2014).

\bibitem{Kit1}
A. Kitaev, {\em{Fault-tolerant quantum computation by anyons}}, Ann. Phys. (N.Y.) 303, 2 (2003).

\bibitem{RH07}
R. Raussendorf and J. Harrington, {\em{Fault-Tolerant Quantum Computation with High Threshold in Two Dimensions}}, Phys. Rev. Lett. \textbf{98}, 190504 (2007).

\bibitem{NC}
M.A. Nielsen and I.L. Chuang, {\em{Quantum Information and Computation}}, Cambridge University Press, 2000.

\bibitem{Veitch2}
Victor Veitch, Seyed Ali Hamed Mousavian, Daniel Gottesman, and Joseph Emerson, {\em{The Resource Theory of Stabilizer Computation}}, New J. Phys. \textbf{16}, 013009 (2014).

\bibitem{Hudson}
 R.L. Hudson, {\em{ When is the Wigner quasi-probability density non-negative?}}, Rep. Math. Phys. 6, 249 (1974).
 
\bibitem{WB}
 Joel J. Wallman, Stephen D. Bartlett, {\em{Nonnegative subtheories and quasiprobability representations of qubits}} Phys. Rev. A \textbf{85}, 062121 (2012).

\bibitem{Howard1}
Mark Howard, Eoin Brennan and Jiri Vala, {\em{Quantum Contextuality with Stabilizer States}}, Entropy \textbf{15}, 2340 (2013).
 
\bibitem{BVN}
Juan Bermejo-Vega, Maarten Van den Nest, {\em{Classical simulations of Abelian-group normalizer circuits with intermediate measurements}}, Int. J Quantum Information and Computation \textbf{14}, 181 (2014).

\bibitem{AB}
J. Anders and D.E. Browne, {\em{Computational Power of Correlations}}, Phys. Rev. Lett. \textbf{102}, 050502 (2009).

\bibitem{Hoban}
M.J. Hoban and D.E. Browne, {\em{Stronger Quantum Correlations with Loophole-Free Postselection}}, Phys. Rev. Lett. \textbf{107}, 120402 (2011).

\bibitem{RR1}
R. Raussendorf, {\em{Contextuality in measurement-based quantum computation}}, Phys. Rev A \textbf{88}, 022322 (2013).

\bibitem{Bombin}
H. Bombin, {\em{Topological Order with a Twist: Ising Anyons from an Abelian Model}}, Phys. Rev. Lett. \textbf{105}, 030403 (2010).

\bibitem{Fowler}
A.G. Fowler, {\em{Low-overhead surface code logical Hada\-mard}}, Quant. Inf. Comp. \textbf{12}, 970 (2012).

\end{thebibliography}
\end{document}